# A Polynomial-Time Algorithm for Pliable Index Coding

Linqi Song and Christina Fragouli

**Abstract**

In pliable index coding, we consider a server with $m$ messages and $n$ clients where each client has as side information a subset of the messages. We seek to minimize the number of broadcast transmissions, so that each client can recover any one unknown message she does not already have. Previous work has shown that the pliable index coding problem is NP-hard and requires at most $\mathcal{O}(\log^2(n))$ broadcast transmissions, which indicates exponential savings over the conventional index coding that requires in the worst case $\mathcal{O}(n)$ transmissions. In this work, building on a decoding criterion that we propose, we first design a deterministic polynomial-time algorithm that can realize the exponential benefits, by achieving, in the worst case, a performance upper bounded by $\mathcal{O}(\log^2(n))$ broadcast transmissions. We extend our algorithm to the $t$-requests case, where each client requires $t$ unknown messages that she does not have, and show that our algorithm requires at most $\mathcal{O}(t\log(n)+\log^2(n))$ broadcast transmissions. We construct lower bound instances that require at least $\Omega(\log(n))$ transmissions for linear pliable index coding and at least $\Omega(t+\log(n))$ transmissions for the $t$-requests case, indicating that both our upper and lower bounds are polynomials of $\log(n)$ and differ within a factor of $\mathcal{O}(\log(n))$. Finally, we provide a probabilistic analysis and show that the required number of transmissions is almost surely $\Theta(\log(n))$, as compared to $\Theta(n/\log(n))$ for index coding. Our numerical experiments show that our algorithm outperforms existing algorithms for pliable index coding by up to $50\%$ less transmissions.

**Index Terms**

Pliable index coding, $t$-requests, polynomial time algorithm, greedy algorithm, random graphs.

## I. Introduction

The conventional index coding problem considers a server with $m$ messages and $n$ clients [3], [4], [5], [6]. Each client has as side-information a subset of the messages and requires a specific

L. Song and C. Fragouli are with the Department of Electrical Engineering, University of California, Los Angeles (UCLA). Email: {songlinqi, christina.fragouli}@ucla.edu.

message she does not have. The aim is to find an efficient way of broadcasting the messages over a noiseless channel such that all clients can be satisfied with the minimum number of transmissions.

Pliable index coding, a variation of the index coding problem first introduced in [7], [8], [9], still considers a server and $n$ clients with side information. However, we now assume that the clients are pliable, and are happy to receive any new message they do not already have. The motivation of pliable index coding stems from the increasing number of emerging applications that transmit types of content. For example, consider a coupon distribution system in a shopping mall and a user who would like to receive coupons. She may have some coupons in her wireless device and may not know exactly the specific coupons that exist, but would be happy to receive any new coupon that she does not have.

For conventional index coding, it has been shown that the problem is NP-hard [10], [11], [12]; we can construct instances that in the worst case may require $\Omega(n)$ transmissions, and even for scalar index coding instances over random graphs, we will almost surely require $\Theta(n/\log(n))$ transmissions [13]. In contrast, pliable index coding can require an exponentially smaller number of transmissions (over index coding), in the worst case $\mathcal{O}(\log^2(n))$ [7], [8]. The result implies that, if we realize that we need to solve a pliable index coding problem as opposed to the conventional index coding problem, we can be exponentially more efficient in terms of the number of transmissions. However, the pliable index coding problem is still NP-hard [7], and thus a natural question is, whether we can efficiently realize these benefits through a polynomial-time algorithm.

The first contribution of this paper is to design a deterministic polynomial-time algorithm that can realize these exponential benefits. We first propose an algebraic decoding criterion for linear pliable index coding, which can be used to determine the validity of a specific linear code for a problem instance. Leveraging this criterion, we design a deterministic polynomial-time algorithm to solve the pliable index coding problem. The algorithm runs in rounds and requires at most $\mathcal{O}(\log^2(n))$ number of transmissions. In each round, we strategically divide the messages into groups and use a greedy transmission scheme to guarantee that a certain fraction of clients are satisfied.

Note that the pliable index coding problem is NP-hard [7] and our proposed approximation algorithm runs in polynomial time. Clearly, our algorithm does not achieve the optimal code length, but we show that it still achieves an upper bound of $\mathcal{O}(\log^2(n))$ in terms of the number of transmissions, which matches the upper bound in [7].

A second contribution is to extend these results to the multiple requests case where each client would like to recover $t$ unkown messages instead of one. In [8], an upper bound of code length $\mathcal{O}(t\log(n)+\log^3(n))$ is shown through probabilistic arguments. We modify our algorithm by introducing weights of clients and messages in the transmission process based on how many new messages a client has received so far. We analytically show that the new algorithm achieves an upper bound of $\mathcal{O}(t\log(n)+\log^2(n))$ for the code length, which is tighter than the upper bound in [8].

Our third contribution is to construct specific instances to provide lower bounds on the required number of transmissions. We construct instances that require $\Omega(\log(n))$ transmissions and $\Omega(t+\log(n))$ transmissions for pliable index coding and the $t$-requests case, respectively. These lower bounds are within a $\mathcal{O}(\log n)$ factor of the upper bounds.

We proceed to provide a probabilistic analysis over random graphs, where the side information sets are populated independently randomly for each client with a certain probability. We show that the required number of transmissions is almost surely $\Theta(\log(n))$, which again is exponentially better than the $\Theta(n/\log(n))$ transmissions required for index coding [13].

Finally, we evaluate the deterministic algorithm performance through numerical experiments. We show that in some cases we can achieve up to $50\%$ savings of transmissions over our previously proposed algorithm.

The paper is organized as follows. Section II describes the pliable index coding formulation and Section III establishes an algebraic criterion for linear pliable index coding; Sections IV and V propose the deterministic polynomial-time algorithms for pliable index coding and its $t$-requests case and analytically prove the upper bound performance of the algorithms; Section VI constructs lower bound instances for the problem; Section VII provides analysis over random instances; Section VIII discusses filed size, connections to minrank, and vector pliable index coding; Section IX discusses related work; Section X carries out numerical experiments; and Section XI concludes the paper.

## II. Problem Formulation

We consider a system with one server and $n$ clients. The server has $m$ messages, represented by symbols in a finite field $b_1, b_2, \dots, b_m \in \mathbf{F}_q$. We will use $[y]$ ($y \in \mathbf{Z}^+$ is a positive integer) to denote the set $\{1, 2, \dots, y\}$ and use $|Y|$ to denote the cardinality of set $Y$ throughout the paper. Each client $i$ has as side information a subset of messages, indexed by $S_i \subseteq [m]$, and requires any new message (or $t$ new messages for $t$-requests case) from the remaining unknown messages, termed request set and indexed by $R_i = [m]\backslash S_i$, where $|R_i| > 0$ (or $|R_i| \geq t$ for $t$-requests case).

Note that unlike index coding where clients are distinguished by their side information sets and requests, in pliable index coding, the clients are only distinguished by their side information sets or request sets. If two clients have the same side information set, a transmission scheme can either satisfy both of them or none of them. Therefore, we assume that no two clients have the same side information set, or equivalently, $S_i \neq S_{i'}$ for all pairs $i, i' \in [n]$, $i \neq i'$. This gives a relationship between $m$ and $n$: $n < 2^m$. Another interesting property of the pliable index coding problem is that when $m > n$, we can always use a transmission scheme for another problem instance with $m \leq n$ to satisfy this problem instance. The observation is that when $m > n$, we can always find a message such that there does not exist any client who only requests this message, according to the pigeon-hole principle. In this case, if we remove this message, then any transmission scheme that satisfies this new problem instance (i.e., by removing this message) can satisfy the original problem instance.

The server first encodes the $m$ original messages into $K$ encoded messages $x_1, x_2, \dots, x_K \in \mathbf{F}_q$ and then makes broadcast transmissions of the encoded messages over a noiseless broadcast channel. Each client receives the broadcasted messages and then decodes them using her side information. We say that a client is *satisfied* if she can successfully recover one new message that she does not already have, or $t$ unknown messages for $t$-requests case, referred to as $t$-*satisfied* or simply *satisfied* when it is clear from the context. Our goal of pliable index coding (or with $t$-requests) is to minimize the total number of transmissions $K$ by designing the encoding and decoding scheme, such that all clients can be satisfied. For ease of exposition, we denote such a problem instance by $(m, n, \{R_i\}_{i\in[n]})$, or $(m, n, \{R_i\}_{i\in[n]}, t)$ for the $t$-requests case.

### A. Encoding and Decoding

Formally, we can express the encoding and decoding processes as follows.

• *Encoding* is represented by an encoding function $f : \mathbf{F}_q^m \to \mathbf{F}_q^K$, where $K$ is the total number of transmissions or code length. The output of the encoding function $(x_1, x_2, \dots, x_K) = f(b_1, b_2, \dots, b_m)$ are the $K$ transmitted messages. We assume that the server has full knowledge of the side information sets for all clients, namely, the server knows $R_i$ for all $i \in [n]$.

• *Decoding*, for client $i \in [n]$, is represented by a decoding function $\phi_i : \mathbf{F}_q^K \times \mathbf{F}_q^{|S_i|} \to \mathbf{F}_q \times [m]$. The output $\phi_i(\{x_k\}_{k\in[K]}, \{b_j\}_{j\in S_i})$ consists of a message in the request set $R_i$ and its index. For the $t$-requests case, the decoding function is $\phi_i^t : \mathbf{F}_q^K \times \mathbf{F}_q^{|S_i|} \to \mathbf{F}_q^t \times [m]^t$. The output $\phi_i^t(\{x_k\}_{k\in[K]}, \{b_j\}_{j\in S_i})$ consists of $t$ messages in the request set $R_i$ and their indices.

We restrict the encoding and decoding schemes to be linear in this paper. In this case, we can further express the encoding and decoding processes as follows.

• *Linear Encoding:* The $k$-th broadcast transmission $x_k$ is a linear combination of $b_1, \dots, b_m$, namely, $x_k = a_{k1}b_1 + a_{k2}b_2 + \dots + a_{km}b_m$, where $a_{kj} \in \mathbf{F}_q$, $j \in [m]$, is the encoding coefficient. Therefore, we can interpret the number of transmissions, $K$, as the *code length* and the $K \times m$ coefficient matrix $\boldsymbol{A}$ with entries $a_{kj}$ as the *coding matrix*. In matrix form, we can write

$$\boldsymbol{x} = \boldsymbol{A}\boldsymbol{b}, \tag{1}$$

where $\boldsymbol{b}$ and $\boldsymbol{x}$ collect the original messages and encoded transmissions, respectively.

• *Linear Decoding:* Given $\boldsymbol{A}$, $\boldsymbol{x}$, and $\{b_j | j \in S_i\}$, the decoding process for client $i$ is to solve the linear equation (1) to get a unique solution of $b_j$ for some $j \in R_i$, or unique solutions $b_{j_1}, b_{j_2}, \dots, b_{j_t}$ for some $j_1, j_2, \dots, j_t \in R_i$ for the $t$-requests case. Clearly, client $i$ can remove her side information messages, i.e., can create $x_k^{(i)} = x_k - \sum_{j\in S_i} a_{kj}b_j$ from the $k$-th encoded transmission. As a result, client $i$ needs to solve the equations

$$\boldsymbol{A}_{R_i}\boldsymbol{b}_{R_i} = \boldsymbol{x}^{(i)}, \tag{2}$$

to retrieve any one message (or $t$ messages) she does not have, where $\boldsymbol{A}_{R_i}$ is the sub-matrix of $\boldsymbol{A}$ with columns indexed by $R_i$; $\boldsymbol{b}_{R_i}$ is the message vector with elements indexed by $R_i$; and $\boldsymbol{x}^{(i)}$ is a $K$-dimensional column vector with element $x_k^{(i)}$.

Our goal is to construct the coding matrix $\boldsymbol{A}$, so that the code length $K$ is minimized.

### *B. Bipartite Graph Representation*

We can represent a pliable index coding problem or its $t$-requests case using an undirected bipartite graph. On one side, a vertex corresponds to a message and on the other side a vertex corresponds to a client. We connect with edges clients to the messages they *do not* have [7], i.e., client $i$ connects to the messages indexed by $R_i$. For instance, in the example in Fig. 1 (a), $R_1 = \{1, 2\}$ and $S_1 = \{3, 4\}$ for client 1; client 4 does not have (and would be happy to receive

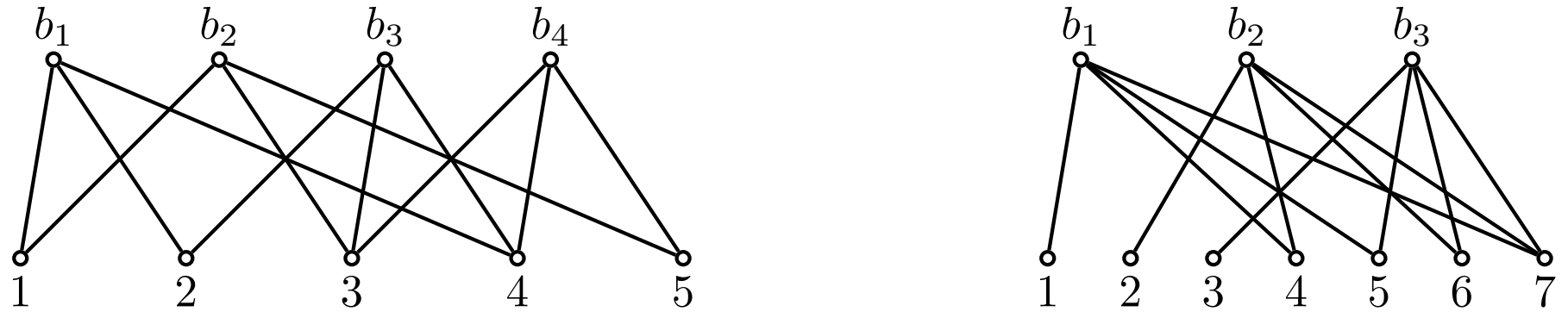

(a) Pliable index coding instance with $m = 4, n = 5$. (b) Pliable index coding instance with $m = 3, n = 7$.

Fig. 1: Illustration of pliable index coding instances.

any of) $b_1$, $b_3$, and $b_4$. In this example, if the server transmits $x_1 = b_1 + b_2$ and $x_2 = b_1 + b_3 + b_4$, then we can write

$$\begin{bmatrix} 1 & 1 & 0 & 0 \\ 1 & 0 & 1 & 1 \end{bmatrix} \begin{bmatrix} b_1 \\ b_2 \\ b_3 \\ b_4 \end{bmatrix} = \begin{bmatrix} x_1 \\ x_2 \end{bmatrix},$$

and the decoding process for client 4 is to solve

$$\begin{bmatrix} 1 & 0 & 0 \\ 1 & 1 & 1 \end{bmatrix} \begin{bmatrix} b_1 \\ b_3 \\ b_4 \end{bmatrix} = \begin{bmatrix} x_1 - b_2 \\ x_2 \end{bmatrix}. \quad (3)$$

## III. An Algebraic Criterion for Linear Pliable Index Coding

We here derive an algebraic criterion that determines whether a client can successfully decode some message given a coding matrix $\boldsymbol{A}$.

Recall that client $i$ needs to solve the linear equations (2) in order to recover a new message. In linear encoding and decoding, (e.g., network coding), we many times solve linear equations to get a unique solution for all elements of the message vector. A key difference in pliable index coding is that, we do not need to identify all the elements of the vector $\boldsymbol{b}_{R_i}$, but only require that any one variable $b_j$, $j \in R_i$ is recovered for client $i$ to be satisfied. Thus we need to achieve a unique solution for one element of the message vector.

We use $\boldsymbol{a}_j$ to denote the $j$-th column of matrix $\boldsymbol{A}$ and $\boldsymbol{A}_{R_i \setminus \{j\}}$ to denote a submatrix of $\boldsymbol{A}$ whose columns are indexed by $R_i$ other than $j$. We also use span$\{\boldsymbol{A}_{R_i \setminus \{j\}}\}$ to denote the linear space spanned by columns of $\boldsymbol{A}$ indexed by $R_i$ other than $j$, i.e., $\{\sum_{l \in R_i \setminus \{j\}} \lambda_l \boldsymbol{a}_l | \lambda_l \in \mathbf{F}_q\}$.

**Lemma 1** (Decoding Criterion)**.** *In a pliable index coding problem $(m, n, \{R_i\}_{i\in[n]})$, given a coding matrix $\boldsymbol{A}$, client $i$ can uniquely decode message $j \in R_i$, if and only if*

$$\boldsymbol{a}_j \notin \textit{span}\{\boldsymbol{A}_{R_i\backslash\{j\}}\}. \tag{4}$$

*Moreover, in the $t$-requests case $(m, n, \{R_i\}_{i\in[n]}, t)$, given a coding matrix $\boldsymbol{A}$, client $i$ can uniquely decode messages $j_1, j_2, \ldots, j_t \in R_i$, if and only if*

$$\boldsymbol{a}_{j_\tau} \notin \textit{span}\{\boldsymbol{A}_{R_i\backslash\{j_\tau\}}\}, \textit{ for all } \tau \in [t]. \tag{5}$$

*Proof.* From linear algebra, we know that the set of solutions for (2) can be expressed as a specific solution plus a vector in the null space of $\boldsymbol{A}_{R_i}$, $\mathcal{N}(\boldsymbol{A}_{R_i})$:

$$\boldsymbol{b}^*_{R_i} + \boldsymbol{b}'_{R_i}, \tag{6}$$

where $\boldsymbol{b}^*_{R_i}$ is a specific solution for (2) and $\boldsymbol{b}'_{R_i}$ is an arbitrary vector in the null space $\mathcal{N}(\boldsymbol{A}_{R_i})$, i.e., $\boldsymbol{b}^*_{R_i}$ is any fixed vector that satisfies $\boldsymbol{A}_{R_i}\boldsymbol{b}^*_{R_i} = \boldsymbol{x}^{(i)}$ and $\boldsymbol{b}'_{R_i}$ is an arbitrary vector that satisfies $\boldsymbol{A}_{R_i}\boldsymbol{b}'_{R_i} = 0$. The requirement that client $i$ can decode message $b_j$ is equivalent to that $b'_j = 0$ for all $\boldsymbol{b}'_{R_i} \in \mathcal{N}(\boldsymbol{A}_{R_i})$, implying that client $i$ can decode message $b_j$ as the unique solution $b^*_j$.

We next argue that $b'_j = 0$ is equivalent to the proposed decoding criterion (4) or (5). We first note that $\boldsymbol{A}_{R_i}\boldsymbol{b}'_{R_i} = b'_j\boldsymbol{a}_j + \sum_{l\in R_i\backslash\{j\}} b_l\boldsymbol{a}_l = 0$ for any vector $\boldsymbol{b}'_{R_i} \in \mathcal{N}(\boldsymbol{A}_{R_i})$.

- *Necessity*. If $\boldsymbol{a}_j \notin \text{span}\{\boldsymbol{A}_{R_i\backslash\{j\}}\}$ is satisfied, then $b'_j = 0$ always holds; otherwise $b'_j\boldsymbol{a}_j + \sum_{l\in R_i\backslash\{j\}} b_l\boldsymbol{a}_l = 0$ for some nonzero $b'_j$ implies that $\boldsymbol{a}_j$ can be expressed as a linear combination of $\boldsymbol{a}_l$, $l \in R_i\backslash\{j\}$, which contradicts the decoding criterion.
- *Sufficiency*. If $b'_j = 0$ always holds, then $\boldsymbol{a}_j \notin \text{span}\{\boldsymbol{A}_{R_i\backslash\{j\}}\}$ is satisfied; otherwise $\boldsymbol{a}_j$ can be expressed as a linear combination of $\boldsymbol{a}_l$, $l \in R_i\backslash\{j\}$ implying that $b'_j = 1$ is also possible, which contradicts the fact that $b'_j = 0$ always holds.

Therefore, we can get a unique solution for $b_j$ if and only if any vector $\boldsymbol{b}'_{R_i}$ in $\mathcal{N}(A_{R_i})$ has a zero value in the element corresponding to $j$. We can then retrieve $b_j$ by any linear equation solving methods for (2). □

For example, considering the instance and coding matrix in Fig. 1 (a) and eq. (3), we have $R_4 = \{1, 3, 4\}$, $\boldsymbol{a}_1 \notin \text{span}\{\boldsymbol{a}_3, \boldsymbol{a}_4\}$, but $\boldsymbol{a}_3 \in \text{span}\{\boldsymbol{a}_1, \boldsymbol{a}_4\}$ and $\boldsymbol{a}_4 \in \text{span}\{\boldsymbol{a}_1, \boldsymbol{a}_3\}$, so client 4 can decode $b_1$ but not $b_3$ and $b_4$. Indeed, client 4 can decode $b_1$ by $b_1 = x_1 - b_2$.

# IV. Binary Field Greedy Algorithm for Pliable Index Coding

In this section, by leveraging the decoding criterion, we design a polynomial-time deterministic algorithm for pliable index coding that achieves a performance guarantee of $\mathcal{O}(\log^2(n))$ in terms of code length. Our algorithm uses operations over the binary field[1] and follows a greedy approach. We first describe our algorithm, which we term BinGreedy, and then show the upper bound performance.

## A. Algorithm Description

Our BinGreedy algorithm is described in Alg. 1. Intuitively, we decide which message we will try to serve to each client: we call *effective clients*, the clients that a specific message aims to satisfy (as we will define more formally in the following), and *effective degree*, the number of such clients each message has. We then create groups of messages that have approximately the same effective degree, and show that because of the regularity of the degree, by coding across only the messages within the group, we can satisfy at least a constant fraction of the effective clients in the group.

The algorithm operates in rounds. Each round has three phases: the sorting phase, the grouping phase and the greedy transmission phase. In the sorting phase, the algorithm sorts the message vertices in a decreasing order in terms of their effective degrees. In the grouping phase, we divide the messages into at most $\log n$ groups based on their effective degrees such that messages in the same group have similar effective degrees. In the transmission phase, to satisfy as many effective clients as possible, the algorithm encodes messages inside a group and makes two transmissions per group, thus in total we use at most $2\log n$ transmissions.

Before giving a detailed description of the algorithm, we first formally introduce the definition of *effective degree* of a message and its *effective clients*.

• *Effective degree* and *effective clients*: given a particular order of the message vertices $\pi = (j_1, j_2, \ldots, j_m)$, the effective degree of message $b_{j_l}$ is defined as the number of $b_{j_l}$'s neighbors who do not connect with message $b_{j'}$, for any $j' = j_1, j_2, \ldots, j_{l-1}$. The neighbors that contribute to $b_{j_l}$'s effective degree are called effective clients of $b_{j_l}$. Let us denote by $N[j]$ the set of neighbors of message $b_j$ and by $N[j_1, j_2, \ldots, j_{l-1}]$ the set $N[j_1] \cup N[j_2] \cup \ldots N[j_{l-1}]$. Formally, the effective clients of message $b_{j_l}$ are defined as $N^\dagger_\pi[j_l] = N[j_l] \backslash N[j_1, j_2, \ldots, j_{l-1}]$ with respect to the order $\pi$. Correspondingly, the effective degree of message $b_{j_l}$ is defined as $d^\dagger_\pi[j_l] = |N^\dagger_\pi[j_l]|$ with respect to $\pi$.

[1] Here, we consider a message $b_j$ to be an element in the finite field $\mathbf{F}_{2^L}$. However, we can see from the construction and analysis of the algorithm that this algorithm works for any fields $\mathbf{F}_q$ with $q \geq 2$.

Note that the effective degree and effective clients for a message $b_j$ may vary when we change the order of the message vertices. We will omit the subscript $\pi$ when it is clear from the context. In our example in Fig. 1 (a), given a message order $b_1, b_2, b_3, b_4$, the effective degrees and clients are $d^\dagger[1] = 3, N^\dagger[1] = \{1, 2, 4\}$, $d^\dagger[2] = 2, N^\dagger[2] = \{3, 5\}$, and $d^\dagger[3] = d^\dagger[4] = 0, N^\dagger[3] = N^\dagger[4] = \emptyset$. Given a different order $b_2, b_4, b_1, b_3$, the effective degrees and clients are $d^\dagger[2] = 3, N^\dagger[2] = \{1, 3, 5\}$, $d^\dagger[4] = 1, N^\dagger[4] = \{4\}$, $d^\dagger[1] = 1, N^\dagger[1] = \{2\}$, and $d^\dagger[3] = 0, N^\dagger[3] = \emptyset$.

In the following, we describe the detailed operations for the three phases in a round.

*1. Sorting Phase*. We sort the messages into a desired order so that the effective degrees of messages are non-increasing. To describe the procedure, we use the bipartite graph representation of pliable index coding (see Section II). We denote by $G$ the original bipartite graph representation of the pliable index coding instance, by $V(G)$ the vertex set of $G$, and by $V(G')$ the vertex set of any subgraph $G'$ of $G$. For a vertex $j \in V(G)$, the set of neighbors of $j$ is denoted by $N[j]$. For a vertex $j$ in an induced subgraph $G'$ of $G$, we define the neighbors of $j$ restricted on subgraph $G'$ as $N_{G'}[j] \triangleq N[j] \cap V(G')$ for all $j \in V(G')$.

• Step 1: We start from the original bipartite graph $G_1 = G$. Find a message vertex $j_1$ with the maximum degree (number of neighbors) in $G_1$, with ties broken arbitrarily. Thus we have $|N_{G_1}[j_1]| \geq |N_{G_1}[j]|$ for all $j \in [m] \backslash \{j_1\}$, where $N_{G_1}[j] = N[j]$.

• Step 2: Consider the subgraph $G_2$ induced by message vertices $[m] \backslash \{j_1\}$ and client vertices $[n] \backslash N[j_1]$. Find a message vertex $j_2$ with maximum degree in the subgraph $G_2$, with ties broken arbitrarily. That is, we have $|N_{G_2}[j_2]| \geq |N_{G_2}[j]|$ for all $j \in [m] \backslash \{j_1, j_2\}$, where $N_{G_2}[j] = N[j] \backslash N[j_1]$.

• Step $l$ ($l = 3, \ldots, m$): Consider the subgraph $G_l$ induced by messages $[m] \backslash \{j_1, j_2, \ldots, j_{l-1}\}$ and clients $[n] \backslash N[j_1, j_2, \ldots, j_{l-1}]$. Find a message vertex $j_l$ with maximum degree in the subgraph $G_l$, with ties broken arbitrarily. That is, we have $|N_{G_l}[j_l]| \geq |N_{G_l}[j]|$ for all $j \in [m] \backslash \{j_1, j_2, \ldots, j_l\}$, where $N_{G_l}[j] = N[j] \backslash N[j_1, j_2, \ldots, j_{l-1}]$.

From the above sorting process, we notice that the effective degrees are $|N[j_1]|$ for message $j_1$, $|N[j_2] \backslash N[j_1]|$ for message $j_2$, ..., $|N[j_l] \backslash N[j_1, j_2, \ldots, j_{l-1}]|$ for $j_l$, etc. It is easy to see that the effective degrees of messages are in a non-increasing order.

*2. Grouping Phase*. We divide the message vertices into $\log(n)$ groups, $\mathcal{M}_1, \mathcal{M}_2, \ldots, \mathcal{M}_{\log(n)}$ based on their effective degrees, such that for message vertex $j \in \mathcal{M}_s$, the effective degree satisfies $n/2^{s-1} \geq d^\dagger[j] > n/2^s$, for $s = 1, 2, \ldots, \log(n)$.

Given the sorting and grouping processes, we have the following property for any message $j$ in group $\mathcal{M}_s$:

$$d^\dagger[j] > n/2^s \triangleq d^{(s)}/2, \text{ and } |N[j] \cap \mathcal{N}_s| \leq n/2^{s-1} \triangleq d^{(s)}, \tag{7}$$

where $\mathcal{N}_s$ is the set of all effective clients of the messages in $\mathcal{M}_s$, namely, $\mathcal{N}_s = \cup_{j' \in \mathcal{M}_s} N^\dagger[j']$. The second part holds because if $|N[j] \cap \mathcal{N}_s| > d^{(s)}$, message $j$, during the sorting and grouping phases, would have effective degree greater than $d^{(s)}$ and would have been assigned in an earlier group (with smaller $s$).

One possible sorting order and grouping for the example in Fig. 1 (a) are: $b_1, b_2, b_3, b_4$ and the only group is $\mathcal{M} = \{1, 2\}$.

*3. Transmission Phase*. We make two transmissions for each message group $\mathcal{M}_s$, using a coding submatrix with 2 rows (one for each transmission). Initially, this submatrix is empty. We sequentially visit each message vertex in $\mathcal{M}_s$ according to the sorting order and create a corresponding column of the coding submatrix, referred to as the coding sub-vector. Hence, a total of $|\mathcal{M}_s|$ steps are carried out for group $\mathcal{M}_s$. The coding sub-vectors are selected from the set $\{(1,0)^T, (0,1)^T, (1,1)^T\}$ such that a maximum number of clients can be satisfied so far. This is explained in detail as follows.

At any step, we say a message $b_j \in \mathcal{M}_s$ is *unvisited* if it is not visited yet up to current step; we say a client $i \in \mathcal{N}_s$ is *unvisited* if it is not connected with the visited messages up to current step. We record the clients that can be satisfied when some message vertices are visited and associated coding sub-vectors are added to the coding submatrix. We denote by $\boldsymbol{A}^{(s)}(j^{(s)})$ a $2 \times |\mathcal{M}_s|$ coding submatrix that consists of the assigned coding sub-vectors corresponding to visited messages and $(0,0)^T$ corresponding to unvisited messages in $\mathcal{M}_s$. In particular, we define the following three types of clients in $\mathcal{N}_s$ when we assign a coding sub-vector from $\{(1,0)^T, (0,1)^T, (1,1)^T\}$ after we visit a message $j^{(s)}$.

• Clients in set $SAT$: we collect in a set $SAT$ the clients in $\mathcal{N}_s$ that can be satisfied by the coding submatrix constructed so far, i.e., $SAT = \{i \in \mathcal{N}_s | i \text{ is satisfied by } \boldsymbol{A}^{(s)}(j^{(s)}).\}$.

• Clients in set $UNSAT$: we collect in a set $UNSAT$ the clients $\mathcal{N}_s$ that are visited and cannot be satisfied by the coding submatrix constructed so far, i.e., $UNSAT = \{i \in \mathcal{N}_s | i$ is visited and not satisfied by $\boldsymbol{A}^{(s)}(j^{(s)})$.}.

• Unvisited clients: obviously, the unvisited clients are only connected to unvisited messages that correspond to columns $(0,0)^T$ in coding submatrix $\boldsymbol{A}^{(s)}(j^{(s)})$. Hence, these unvisited clients are not satisfied by $\boldsymbol{A}^{(s)}(j^{(s)})$.

Note that when a message $j^{(s)}$ is visited at current step, some of the unvisited clients are visited for the first time. These clients are treated as the effective clients of the currently visited message $j^{(s)}$ and added to the set $SAT$, according to the decoding criterion that any vector from $\{(1,0)^T,(0,1)^T,(1,1)^T\}$ is not in the space spanned by some $(0,0)^T$ vectors. After assigning a coding sub-vector from $\{(1,0)^T,(0,1)^T,(1,1)^T\}$ to message $j^{(s)}$, some of the clients previously in $SAT$ may be moved to $UNSAT$. For example, if we assign a coding sub-vector $(1,1)^T$ to message $j^{(s)}$, a client connecting with message $j^{(s)}$ is also connected to $3$ messages that are assigned a vector $(1,0)^T$ and two vectors $(0,1)^T$. According to the decoding criterion, this client is satisfied before current step as $(1,0)^T$ is not in the space spanned by $(0,1)^T$, but is not satisfied after current step as $(1,0)^T$ is in the space spanned by $(0,1)^T$ and $(1,1)^T$ and is the same for $(0,1)^T$ or $(1,1)^T$. Also, some of the clients from $UNSAT$ can be moved to $SAT$. For example, when a client is connecting with 2 visited messages corresponding to the vector $(1,0)^T$, we visit another message that this client is connecting with and assign a coding sub-vector $(0,1)^T$, then this client can be satisfied again.

In the transmission phase, for each step when a message is visited, we assign a coding sub-vector from $\{(1,0)^T,(0,1)^T,(1,1)^T\}$ to maximize the number of currently satisfied clients, i.e., the size of $SAT$. Note that only those clients who are connected with message $j^{(s)}$ will be affected at current step compare with at last step. Maximizing the size of $SAT$ is equivalent to maximizing the number of satisfied clients so far that are connected with $j^{(s)}$ as shown in line $15$ of Alg. $1$.

In our simple example given in Fig. 1 (a), we can construct a coding matrix:

$$\boldsymbol{A}=\begin{bmatrix} 1 & 1 & 0 & 0 \\ 0 & 1 & 0 & 0 \end{bmatrix}.$$

Note that the coding matrix constructed by our algorithm may not be full rank. As such, it suffices to only select a row basis as the coding matrix.

**Algorithm 1** Binary Field Greedy Algorithm (BinGreedy)

```
Initialization: Set 𝒩 = [n].
while 𝒩 ≠ ∅ do
    Sorting and grouping of message vertices:
    Set 𝒩_temp = 𝒩, ℳ_temp = [m].
    for j = 1 : m do
        Find the message j' ∈ ℳ_temp having the maximum number of neighbors in 𝒩_temp, with ties broken arbitrarily.
        Put message j' in the j-th position.
        Remove j' from ℳ_temp and all its neighbors from 𝒩_temp.
    end for
    Group messages into ℳ_1, ℳ_2, ..., ℳ_log(n) message groups based on their effective degrees.
    Greedy transmission:
    for s = 1 : log(n) do
        Initialization: Set 𝒩_s = ∪_{j∈ℳ_s} N†[j] (effective clients neighboring to ℳ_s), SAT = ∅ and UNSAT = ∅.
        for j = 1 : |ℳ_s| do
            Assign a coding sub-vector from {(1,0)^T, (0,1)^T, (1,1)^T} to the j-th message in ℳ_s, denoted by j^(s), aiming at maximizing the number of clients in {i ∈ SAT | i is connected with j^(s)} so far, with ties broken arbitrarily.
            Move from SAT to UNSAT these unsatisfied clients in {i ∈ SAT | i is connected with j^(s)}.
            Add clients in N†[j^(s)] to SAT.
        end for
        Set coding sub-vectors to be (0,0)^T corresponding to messages in groups other than s.
        Remove clients in SAT from 𝒩 and their associated edges.
    end for
end while
```

### B. Example

We now show how the algorithm works through an example. We consider the following problem instance represented by the biadjacency matrix[2] on the left hand side in Fig. 2 (a). In this biadjacency matrix, we have the number of messages $m = 5$, each represented by a column of the matrix, and the number of clients $n = 14$, each represented by a row of the matrix. The request sets are shown as adjacency relationship in the biadjacency matrix, i.e., the $(i, j)$ entry is 1 if and only if client $i$ does not have message $j$.

The sorting and grouping phases are shown in Fig. 2 (a). In the sorting phase, 5 messages are sorted in a non-increasing order according to their effective degrees, as shown on top of the matrix on the right hand side of Fig. 2 (a). We categorize these messages and their associated effective clients into 2 groups, such that the maximum effective degree in a group is not more than twice the minimum effective degree in the group.

Fig. 2 (b) shows the greedy transmission phase for each group. In a group, we sequentially assign coding sub-vectors of length 2 to each message, such that the maximum number of clients

[2] For a bipartite graph $G(U \cup V, E)$, the biadjacency matrix is a $(0, 1)$ matrix of size $|U| \times |V|$, whose $(i, j)$ element equals 1 if and only if $i$ connects $j$.

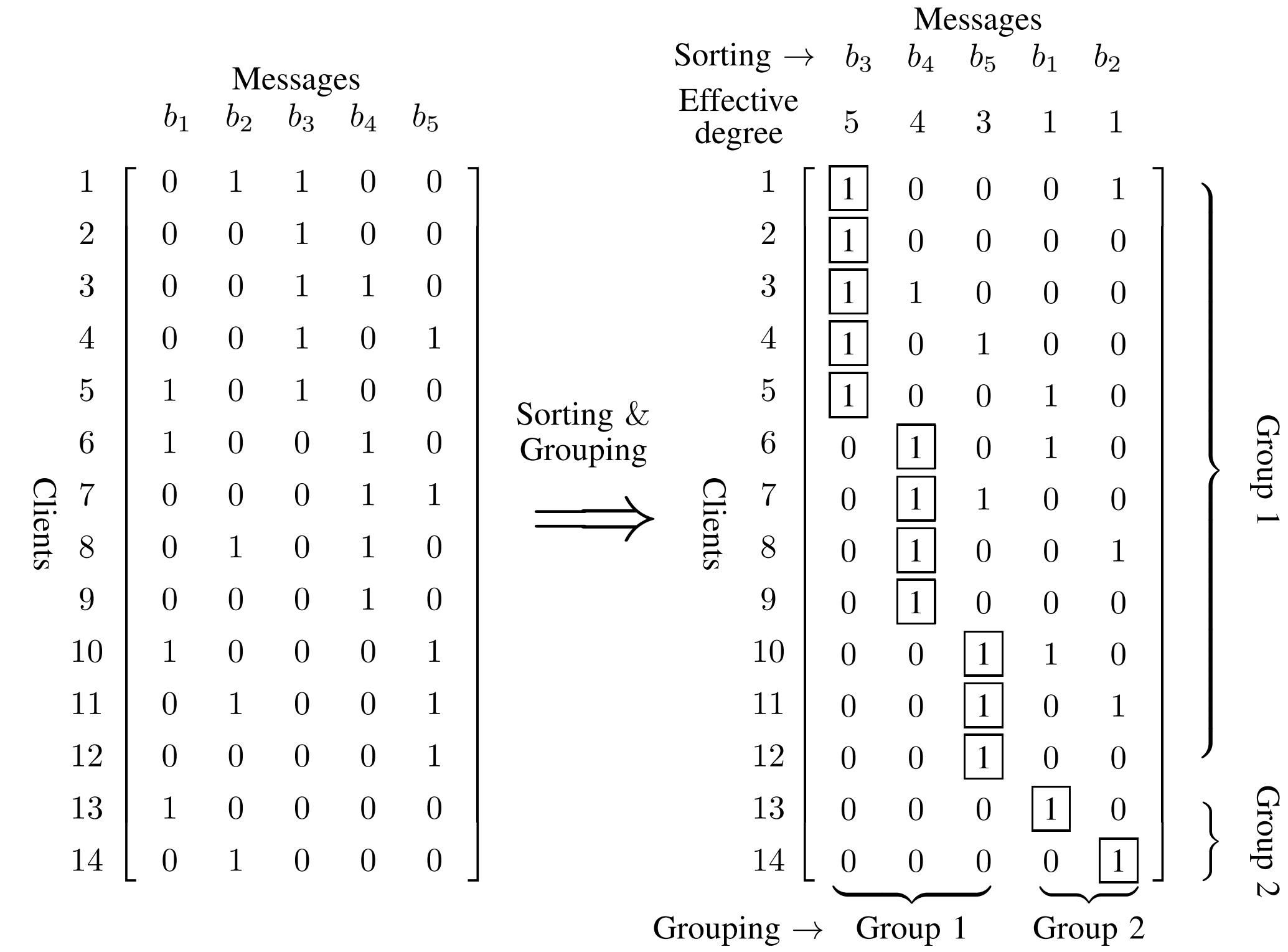

(a) Sorting and grouping phases. The messages are first sorted in a non-increasing order according to their effective degrees, and then are grouped into groups. Clients constributing to effective degree of a message are boxed in the biadjacency matrix.

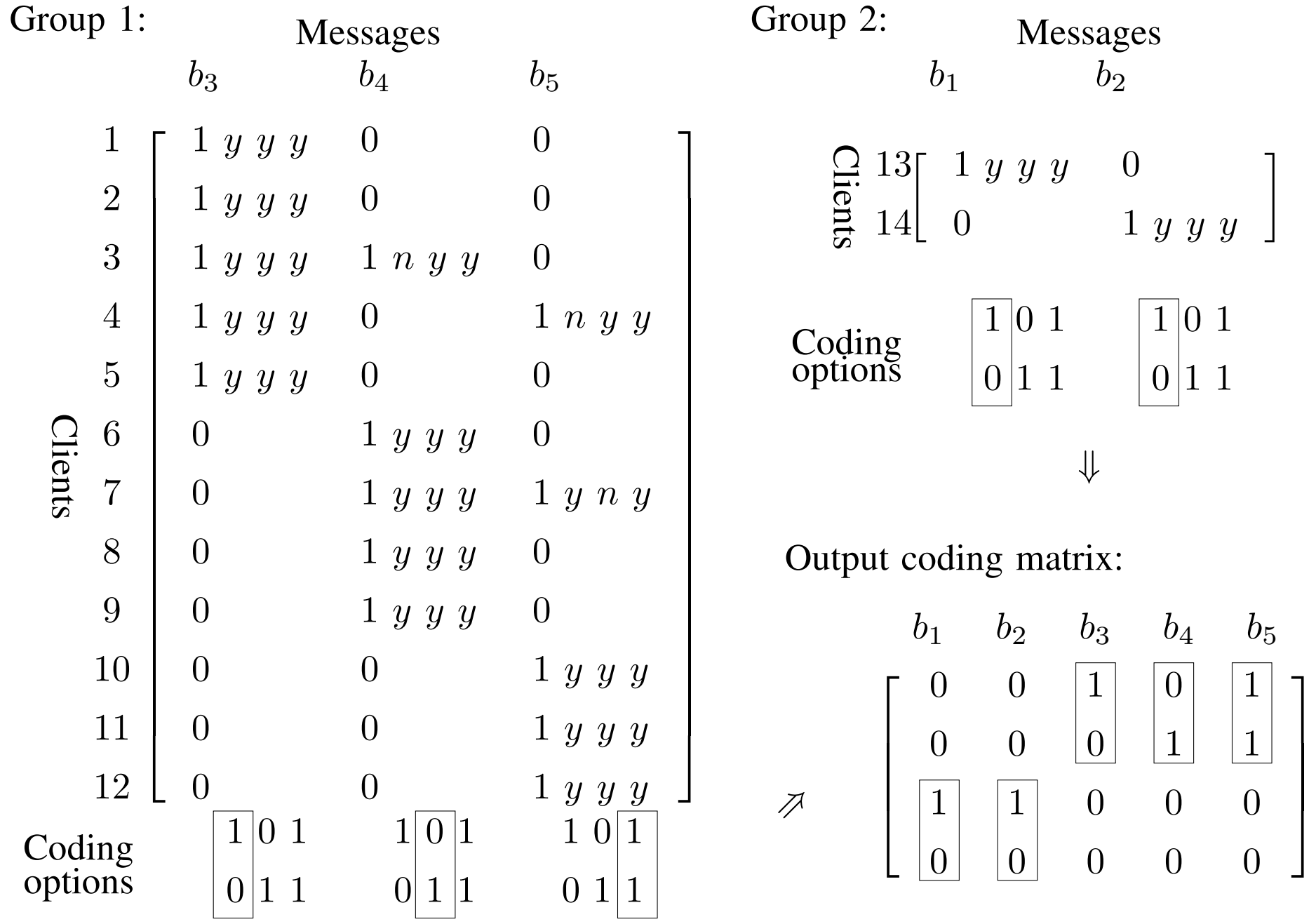

(b) Greedy transmission phase. For each group, coding sub-vectors of length 2 are sequentially assigned to each message, so as to satisfy as many clients as possible at each step. At each step, the coding options are listed to check if a client can be satisfied ($y$) or not ($n$) so far, and selections are boxed.

Fig. 2: An example of running BinGreedy algorithm in 1 round.

can be satisfied so far in the group. For example, in group $1$, when we assign coding sub-vectors for message $b_4$, we find that if $(1, 0)^T$ is chosen, the coding submatrix becomes

$$\boldsymbol{A}^{(1)}(j^{(2)}) = \begin{bmatrix} 1 & 1 & 0 \\ 0 & 0 & 0 \end{bmatrix}, \tag{8}$$

and $4$ clients that are connecting with message $b_4$ can be satisfied up to now (clients $6$-$9$). However, if we $(0, 1)^T$ or $(1, 1)^T$ is chosen, the matrix becomes

$$\boldsymbol{A}^{(1)}(j^{(2)}) = \begin{bmatrix} 1 & 0 & 0 \\ 0 & 1 & 0 \end{bmatrix}, \text{ or } \boldsymbol{A}^{(1)}(j^{(2)}) = \begin{bmatrix} 1 & 1 & 0 \\ 0 & 1 & 0 \end{bmatrix}, \tag{9}$$

and $5$ clients that are connecting with message $b_4$ can be satisfied up to now (clients $3$, $6$-$9$). By comparing these selections, we choose $(0, 1)^T$ (or $(1, 1)^T$) as the coding sub-vector. The final coding matrix achieved by our BinGreedy algorithm is shown at the bottom on the right hand side in Fig. $2$ (b).

Note that for this instance, one round of encoding is enough to satisfy all clients.

### C. Algorithm Performance

To evaluate the worst case performance of our proposed algorithm in terms of the number of transmissions, we first prove the following lemma.

**Lemma 2.** *In Alg. 1, the greedy coding scheme can satisfy at least* $1/3$ *of the effective clients* $\mathcal{N}_s$ *in one round.*

*Proof.* Consider the bipartite subgraph induced by vertices $\mathcal{M}_s \cup \mathcal{N}_s$, i.e., the messages in the group $\mathcal{M}_s$ and their effective clients in $\mathcal{N}_s$. To construct the coding submatrix, at each step we sequentially visit a message vertex $j$ in $\mathcal{M}_s$, following the sorted order, denoted by $1^{(s)}, 2^{(s)}, \ldots, m_s^{(s)} = |\mathcal{M}_s|$, and greedily decide which coding sub-vector will become the $j-th$ column of the coding matrix.

Up to a certain step, a client is either unvisited or satisfied/unsatisfied by the current assignment. Recall that to capture the dynamic changes of satisfied/unsatisfied clients, we define two sets, $SAT$ and $UNSAT$. The first set, $SAT$, collects the clients connecting to messages that have already been visited, and are satisfied by the current assignment of coding sub-vectors according to the criterion in Lemma 1, i.e., for each of these clients, $i$, given the $r$ coding sub-vectors assigned to messages connecting with $i$ and visited by the algorithm so far, $\alpha_1, \alpha_2, \ldots, \alpha_r$, there exists one coding sub-vector $\alpha_{j'}$ $(1 \leq j' \leq r)$ not in the span of the remaining coding sub-vectors: $\alpha_{j'} \notin \text{span}\{\alpha_1, \ldots, \alpha_{j'-1}, \alpha_{j'+1}, \ldots, \alpha_r\}$. The second set, $UNSAT$, collects clients that are associated with messages already visited by the algorithm and cannot be satisfied by current coding -sub-vector assignments. Note that both sets $SAT$ and $UNSAT$ only contain visited clients and there may also exist unvisited clients in $\mathcal{N}_s$.

Initially, both $SAT$ and $UNSAT$ are empty. We gradually add clients from $\mathcal{N}_s$ into these two sets as we go through the messages and assign coding sub-vectors. Our first step is to add all $N^\dagger[1^{(s)}]$ (effective clients of the first message $1^{(s)}$ in $\mathcal{M}_s$) to $SAT$, since any non-zero vector satisfies the decoding criterion for only one message. Recall that at each step, some unvisited clients are visited and become satisfied (being put into the set $SAT$); some satisfied clients may also become unsatisfied (being put into the set $UNSAT$); some unsatisfied clients may become satisfied (being put into the set $SAT$)[3].

We will show that at each step, the number of clients who are moved from $SAT$ to $UNSAT$ is at most $d^{(s)}/3$. Consider the step to assign a vector to message $j$ in $\mathcal{M}_s$. Notice that when we assign a coding sub-vector $(1,0)^T, (0,1)^T$, or $(1,1)^T$ to message $j$, only clients connecting with message $j$ can be affected. We list possibilities for all the $t_0$ clients connected with $j$ and satisfied (in $SAT$) at the beginning of this step:

• Case 1: Assume there are $t_1$ clients who connect with previously visited messages that are assigned one coding sub-vector $(1,0)^T$ and some (perhaps none) coding sub-vectors $(0,1)^T$. In this case, these clients can decode a new message corresponding to the coding sub-vector $(1,0)^T$ since $(1,0)^T$ does not belong in the span of $(0,1)^T$ according to the decoding criterion. Similarly,

• Case 2: $t_2$ clients are satisfied by a $(1,0)^T$, several $(1,1)^T$.

• Case 3: $t_3$ clients are satisfied by a $(0,1)^T$, several $(1,0)^T$.

• Case 4: $t_4$ clients are satisfied by a $(0,1)^T$, several $(1,1)^T$.

• Case 5: $t_5$ clients are satisfied by a $(1,1)^T$, several $(0,1)^T$.

• Case 6: $t_6$ clients are satisfied by a $(1,1)^T$, several $(1,0)^T$.

If we assign a coding sub-vector $(1,0)^T$ to message $j$, the $t_3 + t_6$ clients can still be satisfied according to Lemma 1. Similarly, if we assign a coding sub-vector $(0,1)^T$ or $(1,1)^T$ to message $j$, then the $t_1 + t_5$ or $t_2 + t_4$ clients can still be satisfied.

[3]We can also consider a relaxed scenario that in a round, once a client is put into the set $UNSAT$, she cannot be put into the set $SAT$ again; then the analytical performance of the algorithm described in Theorem 1 does not change based the analysis in this subsection.

Note that $t_1 + t_2 + t_3 + t_4 + t_5 + t_6 \geq t_0$ as there may be overlap among the 6 different cases (e.g., a client is satisfied by one $(1,0)^T$ and one $(0,1)^T$, so she is counted twice in both Case 1 and Case 3). Hence, at least one of $t_3 + t_6$, $t_1 + t_5$, $t_2 + t_4$ should be no less than $t_0/3$; our greedy algorithm will move at most $2t_0/3$ clients from $SAT$ to $UNSAT$. According to the property of our sorting and grouping in eq. (7), the number of message $j$'s neighbors in $\mathcal{N}_s$ who are connected with previously visited messages (and hence are not $j$'s effective clients) is at most $|N[j] \cap \mathcal{N}_s \backslash N^\dagger[j]| \leq d^{(s)} - d^\dagger[j] < d^{(s)}/2$. Moreover, before this step, message $j$'s neighbors in the set $SAT$ must be effective clients of some previously visited messages, implying $t_0 \leq d^{(s)} - d^\dagger[j] < d^{(s)}/2$. Thus, the number of clients being moved from $SAT$ to $UNSAT$ at current step is at most $2t_0/3 < 2d^{(s)}/2/3 = d^{(s)}/3$.

On the other hand, we observe that for message $j$'s effective clients ($j$'s neighbors who are not connected with previously visited messages), any assignment of vectors $(1,0)^T, (0,1)^T$, or $(1,1)^T$ can satisfy them according to the decoding criterion. So, at least $d^\dagger[j] > d^{(s)}/2$ unvisited clients are added to $SAT$. Completing the assignment steps, we can see that at most $2/3$ clients in $\mathcal{N}_s$ cannot be satisfied by this scheme. □

We can now prove the following theorem.

**Theorem 1.** *For the BinGreedy algorithm in Alg. 1, the number of required transmissions is at most* $\frac{2}{log(1.5)} \log^2(n)$.

*Proof.* From Lemma 2, in each round, we have at most $\log(n)$ groups and $2\log(n)$ transmissions such that at least $1/3$ clients are satisfied. This can be repeated for at most $\log(n)/\log(1.5)$ times, where the theorem follows. □

From the construction of our greedy algorithm, we can easily see that the algorithm runs in polynomial time $\mathcal{O}(nm^2 \log(n))$: there are at most $\mathcal{O}(\log(n))$ rounds; for each round, the sorting and grouping phases take time $\mathcal{O}(nm^2)$; and the greedy transmission phase in each round takes time $\mathcal{O}(mn)$.

## V. Binary Field Greedy Algorithm for $t$-requests Case

A straightforward method to solve the $t$-requests case is by repeatedly solving pliable index coding instances $t$ times, resulting in an upper bound $\mathcal{O}(t \log^2(n))$ of the number of broadcast transmissions. In [8] an upper bound of code length $\mathcal{O}(t\log(n)+\log^3(n))$ is proved achievable. In this section, we modify our algorithm to adapt it to the $t$-requests case and prove that this modified algorithm, which we term BinGreedyT, can achieve a tighter upper bound $\mathcal{O}(t\log(n)+\log^2(n))$.

### A. Algorithm Description

The key difference of the BinGreedyT algorithm from the BinGreedy algorithm is the introduction of weights for clients and messages. The main idea behind this is that we would like all clients to receive approximately a similar number of new messages as the transmission proceeds, aiming to avoid that some clients receive too many new messages while others receive too few during the transmission process. For this purpose, we originally assign the same weights for all clients and exponentially reduce the weight of a client each time she can recover a new message. As a result, the algorithm operates in an efficient way which we show to achieve an upper bound $\mathcal{O}(t\log(n)+\log^2(n))$.

We first introduce some new definitions. As the algorithm runs, we say that at some point a client is $\tau$-satisfied if she can decode $\tau$ unknown messages in $R_i$. The ultimate goal of our algorithm is to let all clients to be $t$-satisfied. We again use the bipartite graph representation and denote by $N[j]$ the set of neighbors of message vertex $j$.

• *Weights of clients,* $w_i$*:* we associate a weight $0 \leq w_i \leq 1$ with each of the $n$ clients. Initially, we set $w_i = 1$ for all client $i \in [n]$. This weight will be updated over time as the algorithm is being carried out.

• *Weights of messages,* $w[j]$*:* the weight corresponding to a message vertex $b_j$ is the summation of the weights of $b_j$'s neighbors $N[j]$, i.e., $w[j] = \sum_{i\in N[j]} w_i$.

• *Weights of messages restricted on a subgraph or a client subset,* $w_{G'}[j]$ *or* $w_{\mathcal{N}'}[j]$*:* given a subgraph $G'$ of $G$ (or a subset of clients $\mathcal{N}' \subseteq [n]$), the weight of a message $j$ restricted on the subgraph $G'$ (or on the client subset $\mathcal{N}'$) is the summation of the weights of $b_j$'s neighbors in the subgraph $G'$ (or in the client subset $\mathcal{N}'$), denoted by $w_{G'}[j] = \sum_{i\in N[j]\cap V(G')} w_i$ (or $w_{\mathcal{N}'}[j] = \sum_{i\in N[j]\cap \mathcal{N}'} w_i$).

• *Effective weights and effective neighbors of messages,* $w^\dagger[j]$ *and* $N^\dagger[j]$*:* given a particular order of the message vertices $\pi = (j_1, j_2, \ldots, j_m)$, the effective weight of message $b_{j_l}$ is defined as the sum of weights of $b_{j_l}$'s neighbors who do not connect with message $b_{j'}$, for any $j' = j_1, j_2, \ldots, j_{l-1}$. These neighbors that contribute to $b_{j_l}$'s effective weights are called effective clients of $b_{j_l}$. Let us denote by $N[j_1, j_2, \ldots, j_{l-1}]$ the set of neighbors $N[j_1] \cup N[j_2] \cup \ldots N[j_{l-1}]$.

Formally, the effective clients of message $b_{j_l}$ are defined as $N^\dagger_\pi[j_l] = N[j_l]\backslash N[j_1, j_2, \ldots, j_{l-1}]$ with respect to the order $\pi$. Correspondingly, the effective weights of message $b_{j_l}$ is defined as $w^\dagger_\pi[j_l] = \sum_{i \in N^\dagger_\pi[j_l]} w_i$ with respect to $\pi$. Again, whenever the order $\pi$ is clear from the context, we will omit the subscripts of the effective weights and effective neighbors.

We next describe the algorithm in Alg. 2. The algorithm operates again in rounds and each round has the same three phases. There are two differences here: one is that the sorting, grouping, and transmissions are based on the effective weights of messages, instead of effective degrees; the other is that we have messages categorized into $\log(n) + 1$ groups, with an additional group to gather messages with "small" weights and we do not encode messages within this group when making transmissions. In every $2\log(n)$ transmissions, we want to make sure that clients worth a certain fraction of weight can decode a new message, such that the total weight of clients in the system is decreasing at a fixed ratio during $2\log(n)$ transmissions. This will result in the claimed performance.

*1. Sorting Phase*. In the sorting phase, we use a similar technique as in Section IV to sort the messages into a non-increasing order according to their effective weights instead of their effective degree.

*2. Grouping Phase*. Let us denote by $W = w[1]$ the maximum weight of the messages. We divide the message vertices into $\log(n) + 1$ groups, $\mathcal{M}_1, \mathcal{M}_2, \ldots, \mathcal{M}_{\log(n)}, \bar{\mathcal{M}}$ based on their effective weights with respect to the above order. For the first $\log(n)$ groups, a message vertex $j$ is in groups $\mathcal{M}_s$, if and only if the effective weights satisfies $W/2^{s-1} \geq w^\dagger[j] > W/2^s$. For the remaining messages with "small" weights, i.e., no more than $W/n$, we put them into the last group $\bar{\mathcal{M}}$. We say a client $i$ in group $s$ if it contributes to the effective weight of a message vertex in group $s$. The set of clients in group $s$ is denoted by $\mathcal{N}_s$.

According to the above sorting and grouping processes, we have the following property for the message $j$ in group $\mathcal{M}_s$ ($s = 1, 2, \ldots, \log(n)$):

$$w^\dagger[j] > W/2^s, \text{ and } \sum_{i \in N[j] \cap \mathcal{N}_s} w_i \leq W/2^{s-1}, \tag{10}$$

where $\mathcal{N}_s = \cup_{j' \in \mathcal{M}_s} N^\dagger[j']$. The summation term $\sum_{i \in N[j] \cap \mathcal{N}_s} w_i$ in the second part can be seen as the "weight of message $j$ restricted on client set $\mathcal{N}_s$". This holds because otherwise the message $j$ will be assigned in a group less than $s$ in the sorting and grouping phases.

*3. Transmission Phase*. In the transmission phase, we ignore the last group $\bar{\mathcal{M}}$ and make two

transmissions for each message group $\mathcal{M}_s$ ($s = 1, 2, \ldots, \log(n)$), using a coding submatrix with 2 rows (one for each transmission). We sequentially create this submatrix by visiting each of the messages in group $\mathcal{M}_s$, according to the sorting order, and adding for each message one column to the coding submatrix (we refer to this column as the coding sub-vector associated with this message). So in total we have $|\mathcal{M}_s|$ steps for group $\mathcal{M}_s$. At each step, we select each coding sub-vector to be one in the set $\{(1,0)^T, (0,1)^T, (1,1)^T\}$, such that it can satisfy the maximum weight of clients in $\mathcal{N}_s$ up to the current step.

**Algorithm 2** Binary Field Greedy Algorithm for $t$-requests (BinGreedyT)

**Initialization**: Set $\mathcal{N} = [n]$, $w_i = 1$ for all $i \in [n]$.
**while** $\mathcal{N} \neq \emptyset$ **do**
  **Sorting**:
  Set $\mathcal{N}_{temp} = \mathcal{N}, \mathcal{M}_{temp} = [m]$.
  **for** $j = 1 : m$ **do**
    Find the message $j' \in \mathcal{M}_{temp}$ having the maximum weight of neighbors in $\mathcal{N}_{temp}$, i.e., $j' = \arg\max_{j'' \in \mathcal{M}_{temp}} \sum_{i \in \mathcal{N}_{temp} \cup N[j'']} w_i$, with ties broken arbitrarily.
    Put message $j'$ in the $j$-th position.
    Remove $j'$ from $\mathcal{M}_{temp}$ and all its neighbors from $\mathcal{N}_{temp}$.
  **end for**
  **Grouping**: Group messages into $\log(n) + 1$ groups based on their effective weights.
  Set $W = w[1]$.
  For the first $\log(n)$ groups, put messages whose effective weights are between $W/2^s$ and $W/2^{(s-1)}$ into group $s$. Put the remaining messages into the last group. Then we have the groups: $\mathcal{M}_1, \mathcal{M}_2, \ldots, \mathcal{M}_{\log(n)}, \bar{\mathcal{M}}$
  **Greedy transmission**:
  **for** $s = 1 : \log(n)$ **do**
    **Initialization**: Set $\mathcal{N}_s = \cup_{j \in \mathcal{M}_s} N^\dagger[j]$ (effective clients neighboring to $\mathcal{M}_s$), $SAT = \emptyset$ and $UNSAT = \emptyset$.
    **for** $j = 1 : |\mathcal{M}_s|$ **do**
      Assign a coding sub-vector from $\{(1,0)^T, (0,1)^T, (1,1)^T\}$ to the $j$-th message in $\mathcal{M}_s$, denoted by $j^{(s)}$, aiming at maximizing the weight of clients in $\{i \in SAT | i \text{ is connected with } j^{(s)}\}$ so far, with ties broken arbitrarily.
      Move from $SAT$ to $UNSAT$ these unsatisfied clients in $\{i \in SAT | i \text{ is connected with } j^{(s)}\}$.
      Add clients in $N^\dagger[j^{(s)}]$ to $SAT$.
    **end for**
    Set coding sub-vectors to be $(0,0)^T$ corresponding to messages in groups other than $s$.
    Update clients' weights in $SAT$: $w_i = w_i/2$ for all $i \in SAT$; add one message $b_{j'}$ that $i$ can decode to her side information set $S_i$ and remove this edge between $i$ and $b_{j'}$.
    For all $i \in SAT$, check if $i$ is $t$-satisfied:
    **if** $w_i = 1/2^t$ **then**
      Remove client $i$ from $\mathcal{N}$ and the associated edges.
    **end if**
  **end for**
**end while**

After a round of at most $2\log(n)$ transmissions, if a client $i$ can decode one new message, we reduce the weight by a half: $w_i \to \frac{w_i}{2}$, and add one of her decoded messages in the side-information set $S_i$. If this weight equals to $1/2^t$, i.e., client $i$ is $t$-satisfied, we remove this vertex and its associated edges from the graph. We repeat the process until all clients are $t$-satisfied.

From the described procedure, it follows that the BinGreedyT algorithm reduces to the BinGreedy algorithm for the $t = 1$ case.

### B. Algorithm Performance

We aim to show that the above described algorithm has a performance guarantee upper bounded by $\mathcal{O}(t\log(n) + \log^2(n))$. We first show that in each round, after the $\mathcal{O}(\log(n))$ transmissions, the total weight of clients is at most $\frac{11}{12}$ of that before the $\mathcal{O}(\log(n))$ transmissions, denoted by $W_T$. This implies that the weight is exponentially decreasing, hence, as shown later, we can argue that at most $\mathcal{O}(\log(n) + t)$ rounds are needed.

**Lemma 3.** *The sum of clients' weights in the first* $\log(n)$ *groups is at least* $1/2$ *of the total weight* $W_T$*, i.e.,* $\sum_{s=1}^{\log(n)} \sum_{i\in\mathcal{N}_s} w_i \geq W_T/2$.

*Proof.* To see this, recall that the maximum weight of a message is $W$. According to our sorting and grouping phases in Alg. 2, after $\log(n)$ groups, the maximum weight of a message is at most $W/2^{\log(n)} = W/n$. Then the total weight of clients in group $\bar{\mathcal{M}}$ is at most $\frac{W}{n}n = W$. This means that the sum of clients' weights in the first $\log(n)$ groups is at least $W_T/2$. □

Consider the subgraph induced by vertices $\mathcal{M}_s \cup \mathcal{N}_s$ corresponding to the transmissions of group $\mathcal{M}_s$ ($s = 1, 2, \log(n)$) in a certain round. Similar to Alg. 1, at each step, we sequentially assign to a message a coding sub-vector. We say a client $i \in \mathcal{N}_s$ is *unvisited* if it is not connected with the visited messages up to current step. We introduce two sets $SAT$ and $UNSAT$ to dynamically evaluate whether each client is satisfied or not up to the current step (by only considering messages visited up to now and disregarding all unvisited messages), so as to satisfy the maximum weight of clients up to now. Assume the effective weight of a message $j$ in $\mathcal{M}_s$ is between $w^{(s)}/2$ and $w^{(s)}$. Using the property described in eq. (10), with the same technique as in Section IV, we can show that at each step the weight of clients who are moved from $SAT$ to $UNSAT$ is at most $w^{(s)}/3$. (For completeness, we put the proof of this claim in Appendix A.)

On the other hand, we observe that for message $j$'s effective clients ($j$'s neighbors who are not connected with previously visited messages), any assignment of vectors $(1, 0)^T, (0, 1)^T$, or $(1, 1)^T$ can satisfy them once according to the decoding criterion. Hence, at least $w^\dagger[j] > w^{(s)}/2$ worth of unvisited new clients are added to $SAT$.

Completing the assignment steps, we can see that clients worth at most $2/3$ weight in $\mathcal{N}_s$ cannot be satisfied by this coding scheme. Therefore, in a round, clients who can decode one new message count for at least $\frac{1}{3}\frac{1}{2} = \frac{1}{6}$ the total weight $W_T$. According to our weight updating rule that the weights of these clients will be reduced by at least a half: $w_i \to \frac{w_i}{2}$ (or $w_i \to 0$ if $t$-satisfied), resulting in a $\frac{1}{12}$ weight decreasing in total. Or equivalently, the total weight after one round is at most $\frac{11W_T}{12}$.

Therefore, we have the following theorem.

**Theorem 2.** *For the BinGreedyT algorithm in Alg. 2, the number of required transmissions is at most* $\mathcal{O}(t\log(n) + \log^2(n))$.

*Proof.* From the above argument, after each round, we have at most $2\log(n)$ transmissions such that the remaining weight becomes at most $11/12$ of that before this round. Initially, the total weight is $n$. Hence, after $\mathcal{O}(t+\log(n))$ rounds, the total weight is no more than $n(\frac{11}{12})^{\mathcal{O}(t+\log(n))} \leq 1/2^t$. Since the weight for a client who is not $t$-satisfied is at least $1/2^{t-1} > 1/2^t$, all clients are $t$-satisfied after $\mathcal{O}(t+\log(n))$ rounds of transmissions. The upper bound is proved. □

Note that the time complexity of this algorithm is bounded by $\mathcal{O}((t+\log(n))nm^2)$: we need at most $\mathcal{O}(t+\log(n))$ rounds in the algorithm, and in each round, the algorithm takes $\mathcal{O}(nm^2)$ time to perform sorting and grouping and takes $\mathcal{O}(nm)$ time to perform greedy transmission.

## VI. Lower Bounds

In this section, we provide instances for pliable index coding and $t$-requests case that require at least $\Omega(\log(n))$ and $\Omega(t+\log(n))$ transmissions, respectively.

### *A. A Lower Bound for Pliable Index Coding*

To show a lower bound, we consider the following pliable index coding instances that we term *complete instances*, and define as follows. In a complete instance, we have $n = 2^m - 1$. The requirement set $R_i$ is the $i$-th element of the set $2^{[m]} \backslash \emptyset$, where $2^{[m]}$ is the power set of $[m]$. An example of the complete instance with $m = 3$ is shown in Fig. 1 (b).

**Theorem 3.** *In a complete instance* $(m, n, \{R_i\}_{i\in[n]})$*, the optimal number of transmissions is* $\Omega(\log(n))$.

*Proof.* Obviously, we can trivially satisfy all clients with $m = \log(n)$ transmissions, where each $b_j$ is sequentially transmitted once. We argue that we cannot do better by using induction. We will prove that the rank of the coding matrix $\boldsymbol{A}$ needs to be at least $m$ for the clients to be satisfied according to Lemma 1. Let $J$ denote a subset of message indices; for the complete instance, Lemma 1 needs to hold for any subset $J \subseteq [m]$.

• For $|J| = 1$, to satisfy the clients who miss only one message, no column of the coding matrix $\boldsymbol{A}$ can be zero. Otherwise, if for example, column $j_1$ is zero, then the client who only requests message $b_{j_1}$ cannot be satisfied. So $\text{rank}(\boldsymbol{A}_J) = 1$ for $|J| = 1$.

• Similarly, for $|J| = 2$, any two columns of the coding matrix must be linearly independent. Otherwise, if for example, columns $j_1$ and $j_2$ are linearly dependent, then $\boldsymbol{a}_{j_1} \in \text{span}\{\boldsymbol{a}_{j_2}\}$ and $\boldsymbol{a}_{j_2} \in \text{span}\{\boldsymbol{a}_{j_1}\}$, and the clients who only miss messages $b_{j_1}$ and $b_{j_2}$ cannot be satisfied. So $\text{rank}(\boldsymbol{A}_J) = 2$.

• Suppose we have $\text{rank}(\boldsymbol{A}_J) = l$ for $|J| = l$. For $|J| = l + 1$, we can see that if all clients who only miss $l+1$ messages can be satisfied, then for some $j \in J$, we have $\boldsymbol{a}_j \notin \text{span}\{\boldsymbol{A}_{J\setminus\{j\}}\}$ according to Lemma 1. Therefore, $\text{rank}(\boldsymbol{A}_J) = \text{rank}(\boldsymbol{a}_j) + \text{rank}(\boldsymbol{A}_{J\setminus\{j\}}) = 1 + l$.

Therefore, to satisfy all the clients, the rank of the coding matrix $\boldsymbol{A}$ is $m$, resulting in $K \geq m$, from which the result follows. □

From Theorem 3, we have two observations: 1) We note that the upper bound is $\mathcal{O}(\log^2(n))$ and the lower bound is $\Omega(\log(n))$, which shows that the upper and lower bounds are poly-logarithmic in $n$ (i.e., in the order of polynomial of $\log(n)$) and differ in a factor of $\mathcal{O}(\log(n))$; 2) If we apply our binGreedy algorithm for the complete instance, we achieve a code length of $\log(n)$ as well, since we can divide the messages into $\log(n)$ groups, each consisting of one message.

### *B. A Lower Bound for $t$-requests Case*

We again use complete instances to derive a lower bound for the $t$-requests case. Note that the complete instance for $t$-requests case needs to satisfy $|R_i| \geq t$ for all $i \in [n]$, so we add $t - 1$ dummy messages to the complete instance for $t = 1$ case. Using the bipartite graph representation, the complete instance for $t$-requests is as follows. There are $m$ messages and $n = 2^{m-t+1} - 1$ clients. We divide the messages into 2 types. The first $\log(n + 1) \triangleq m_1$ messages are the Type-1 messages and the remaining $t - 1 \triangleq m_2$ messages are the Type-2 messages. All $n$ clients are connected with all the Type-2 messages. We denote by $J_i \subseteq [m_1]$ the set of Type-1 messages a client $i$ is connected to, i.e., $J_i$ is the set of indices of Type-1 messages that $i$ requires. Each $J_i$ of the $n$ clients is a unique subset of $[m_1]$ except the empty set. Note that there are in total $2^{m_1} - 1 = n$ such unique subsets.

**Theorem 4.** *In a complete instance $(m, n, \{R_i\}_{i\in[n]}, t)$, the optimal number of transmissions is $\Omega(t + \log(n))$.*

*Proof.* Clearly, $m = \log(n+1) + t - 1$ transmissions are enough to satisfy all clients. So we only show at least $\Omega(t + \log(n))$ transmissions are needed.

By abuse of notation, let us denote by $1^{(1)}, 2^{(1)}, \ldots, m_1^{(1)}$ and $1^{(2)}, 2^{(2)}, \ldots, m_2^{(2)}$ the indices of Type-1 and Type-2 messages and by $[1^{(1)} : m_1^{(1)}]$ and $[1^{(2)} : m_2^{(2)}]$ the sets of these two types of messages.

Suppose the coding matrix for a $t$-requests case problem is $\mathbf{A}$. We denote by $\mathbf{A}_{J\cup[1^{(2)}:m_2^{(2)}]}$ the submatrix of $\mathbf{A}$ consisting of columns indexed by $J \cup [1^{(2)} : m_2^{(2)}]$, where $J \subseteq [1^{(1)} : m_1^{(1)}]$ is a subset of indices of Type-1 messages. We will use induction to prove that the rank of the coding matrix $\mathbf{A}$ needs to be at least $m$ for all the clients to be $t$-satisfied according to the decoding criterion. In the complete instance, the decoding criterion needs to hold for all clients, or for all $|J| = 1, 2, \ldots, m_1$.

For $J \subseteq [1^{(1)} : m_1^{(1)}]$ and $|J| = 1$, i.e., to satisfy the clients who miss only one Type-1 message, we need $\text{rank}(\boldsymbol{A}_{J\cup[1^{(2)}:m_2^{(2)}]}) = t$. Since otherwise, for example if the only column $j_1 \in [1^{(1)} : m_1^{(1)}]$ and all $t-1$ columns in $[1^{(2)} : m_2^{(2)}]$ are not linearly independent, then the clients who requests messages $\{j_1\} \cup [1^{(2)} : m_2^{(2)}]$ cannot be $t$-satisfied according to the decoding criterion. So $\text{rank}(\boldsymbol{A}_{J\cup[1^{(2)}:m_2^{(2)}]}) = t$ for all $|J| = 1$.

Assume we have $\text{rank}(\boldsymbol{A}_{J\cup[1^{(2)}:m_2^{(2)}]}) = l + t - 1$ for all $J \subseteq [1^{(1)} : m_1^{(1)}]$ with $|J| = l$. For $J \subseteq [1^{(1)} : m_1^{(1)}]$ and $|J| = l+1$, we can see that according to the induction hypothesis, $\text{rank}(\boldsymbol{A}_{J\cup[1^{(2)}:m_2^{(2)}]}) \geq l + t - 1$. If $\text{rank}(\boldsymbol{A}_{J\cup[1^{(2)}:m_2^{(2)}]}) = l + t - 1$, then for any column $j \in J$, $\boldsymbol{a}_j \in \text{span}\{\boldsymbol{A}_{J\cup[1^{(2)}:m_2^{(2)}]\backslash\{j\}}\}$, since columns in $J \cup [1^{(2)} : m_2^{(2)}]\backslash\{j\}$ consist of a basis for this submatrix from the induction hypothesis. Hence, $b_j$ (for any $j \in J$) cannot be decoded by the client who is only connected with $J \cup [1^{(2)} : m_2^{(2)}]$. This client can decode at most $t-1$ messages and cannot be $t$-satisfied. As a result, $\text{rank}(\boldsymbol{A}_{J\cup[1^{(2)}:m_2^{(2)}]}) = l + t$, from which the result follows. □

# VII. Pliable Index Coding Over Random Graphs

We use a bipartite graph described in Section II to represent a problem instance. Here, we consider the random problem instance represented by a random bipartite graph $B(m, n, p)$ [14], where there are $m$ messages and $n$ clients, and there is an edge between client $i$ and message $j$ with probability $p$, i.e., $\Pr\{j \in R_i\} = p$. We aim to calculate the "average" code length, or with high probability what is the required number $K$ of transmissions. We say that a random problem instance $B(m, n, p)$ *almost surely* needs a code length of $K(m, n, p)$ if the probability that the code length is $K(m, n, p)$ tends to 1 as $m$ and $n$ tend to infinity. Next, we show that a random graph $B(m, n, p)$ *almost surely* requires a code length of $\Theta(\log(n))$.

## A. Lower Bound

To prove a lower bound on $K$, we introduce the concept of a *coding structure*, which is a collection of $K \times m$ matrices $\boldsymbol{A}$ with elements in a finite field $\mathbf{F}_q$ that satisfy a set of properties. Formally, a *coding structure* $S(J^{(1)}, J^{(2)}, J^{(3)})$, or shortly $S$, is defined as $S(J^{(1)}, J^{(2)}, J^{(3)}) \triangleq \{\boldsymbol{A} \in \mathbf{F}_q^{K\times m} | \boldsymbol{A}$ satisfies Properties (1) (2) (3)$\}$, where $J^{(1)}, J^{(2)}, J^{(3)} \subseteq [m]$ are disjoint subsets of message indices, $|J^{(1)}| + |J^{(2)}| = K$, $|J^{(2)}| = |J^{(3)}|$, and the properties are listed as follows.

**Property.**

*(1) Column vectors indexed by $J^{(1)}$ and $J^{(2)}$ contain a column basis of matrix $\boldsymbol{A}$.*

*(2) For any column $j' \in J^{(1)}$, the corresponding column vector is not in the linear space spanned by other column vectors of matrix $\boldsymbol{A}$, i.e., $\boldsymbol{a}_{j'} \notin span\{\boldsymbol{a}_j | j \in [m]\backslash\{j'\}\}$.*

*(3) For any column $j'' \in J^{(2)} \cup J^{(3)}$, the corresponding column vector is in the linear space spanned by other column vectors indexed by $J^{(2)} \cup J^{(3)}$, i.e., $\boldsymbol{a}_{j''} \in span\{\boldsymbol{a}_j | j \in J^{(2)} \cup J^{(3)} \backslash \{j''\}\}$.*

Consider a specific $K \times m$ coding matrix $\boldsymbol{A}$. We next describe a procedure that maps the matrix $\boldsymbol{A}$ (in a non-unique way) to some coding structure $S(J^{(1)}, J^{(2)}, J^{(3)})$. Additionally, given three disjoint subsets of message indices $J^{(1)}, J^{(2)}, J^{(3)} \subseteq [m]$, $|J^{(1)}| + |J^{(2)}| = K$, $|J^{(2)}| = |J^{(3)}|$, we can easily find some matrix $\boldsymbol{A}$ that satisfies Properties (1), (2), and (3). For example, we can construct a matrix $\boldsymbol{A}$, where the submatrix consisting of the first $|J^{(1)}|$ rows and columns indexed by $J^{(1)}$ is an identity matrix; the submatrix consisting of the last $|J^{(2)}|$ rows and columns indexed by $J^{(2)}$ is an identity matrix; and the submatrix consisting of the last $|J^{(2)}|$ rows and columns indexed by $J^{(3)}$ is an identity matrix. Thus, if we denote the set of all $K \times m$ coding matrices by $\mathcal{A}$ and the union of all coding structures by $\mathcal{S} = \cup S(J^{(1)}, J^{(2)}, J^{(3)})$, it is easy to see that $\mathcal{S} = \mathcal{A}$.

*Mapping Procedure:* In the following, we will call the columns in $J^{(1)}$, $J^{(2)}$ and $J^{(3)}$ Type-1, Type-2 and Type-3 columns, respectively. We will use the notation $K_1 = |J^{(1)}|$, $K_2 = |J^{(2)}|$, $K_3 = |J^{(3)}|$. We will show that we can select $J^{(1)}$, $J^{(2)}$ and $J^{(3)}$ so that $K_1 + K_2 = K$, $K_2 = K_3$ and properties (1)-(3) are satisfied. Note that a matrix $\boldsymbol{A}$ could be mapped to multiple structures, since there may exist different choices for selecting the columns in $J^{(1)}$, $J^{(2)}$ and $J^{(3)}$.

• In the coding matrix $\boldsymbol{A}$, find an arbitrary column basis, i.e., a maximum number of linearly independent column vectors. There are at most $K$ such columns and without loss of generality, we assume these columns are indexed by $1, 2, \ldots, K'$, where $K' \leq K$.

• We categorize all $m$ column vectors into 3 groups: 2 groups for these $K'$ basis column vectors and a third group for the remaining $m - K'$ column vectors.

– Group 1: $\boldsymbol{A}^{(1)} = \{\boldsymbol{a}_{j_1} | j_1 \in [K'], \boldsymbol{a}_{j_1} \notin \mathit{span}\{\boldsymbol{a}_j | j \in [m] \backslash \{j_1\}\}\}$. Group 1 consists of column vectors that are not in the linear space spanned by all other column vectors of matrix $\boldsymbol{A}$. We assume $K_1 \leq K'$ such vectors, and without loss of generality, we assume these vectors in Group 1 are indexed by $1, 2, \ldots, K_1$. These are the Type-1 columns.

– Group 2: $\boldsymbol{A}^{(2)} = \{\boldsymbol{a}_{j_2} | j_2 \in [K'], \boldsymbol{a}_{j_2} \in \mathit{span}\{\boldsymbol{a}_j | j \in [m] \backslash \{j_2\}\}\}$. Group 2 consists of column vectors that are in the linear space spanned by all other column vectors of matrix $\boldsymbol{A}$. We assume $K_2 = K' - K_1$ such vectors, and without loss of generality, we assume these vectors in Group 2 are indexed by $K_1 + 1, K_1 + 2, \ldots, K_1 + K_2 = K'$. These are the Type-2 columns.

– Group 3: $\boldsymbol{A}^{(3)} = \{\boldsymbol{a}_{j_3} | j_3 \notin [K']\}$. Group 3 consists of the remaining $m - K'$ column vectors.

• We select and label $K_3$ columns in Group 3 as Type-3 columns as follows. We consider the submatrix of $\boldsymbol{A}$ from removing all $K_1$ columns in Group 1. Initially, we mark all $K_2$ columns in Group 2 as *active* and we will repeatedly *deactivate* them in the following steps.

1) We pick an arbitrary non-zero vector $\boldsymbol{a}_j$ from Group 3.

2) Label vectors or discard them according to the following rule. We observe that after removing the first $K_1$ columns, the $K_2$ column vectors in Group 2 are a basis for the remaining $K \times (m - K_1)$ submatrix. Then the vector $\boldsymbol{a}_j$ that is picked up in Step 1) can be uniquely represented by a linear combination of these basis vectors in Group 2, i.e., $\boldsymbol{a}_j = \lambda_{K_1+1}\boldsymbol{a}_{K_1+1} + \lambda_{K_1+2}\boldsymbol{a}_{K_1+2} + \ldots + \lambda_{K_1+K_2}\boldsymbol{a}_{K_1+K_2}$. Here we can consider $(\lambda_{K_1+1}, \lambda_{K_1+2}, \ldots, \lambda_{K_1+K_2})$ as co-ordinates under this basis.

Using this linear expansion for $\boldsymbol{a}_j$, we consider the basis vectors in Group 2 that correspond to the non-zero coordinates, i.e., $\boldsymbol{A}^{*(2)} = \{\boldsymbol{a}_{j_2} \in \boldsymbol{A}^{(2)} | \lambda_{j_2} \neq 0\}$. If no vectors in $\boldsymbol{A}^{*(2)}$ are marked *active*, then remove column $j$ without labeling it. If any of these basis vectors is marked *active*, then label the column $j$ as Type-3 column, remove it, and mark all column vectors in $\boldsymbol{A}^{*(2)}$ as *inactive* if they are still *active*.

3) Repeat Steps 1) and 2) until all vectors in Group 2 are marked *inactive*. This can always be achieved. Indeed, according to the definition of Group 2, any column vector $\boldsymbol{a}_{j_2} \in \boldsymbol{A}^{(2)}$ can be represented as a linear combination of the other column vectors of matrix $\boldsymbol{A}$. So, $\boldsymbol{a}_{j_2}$ always appears as a non-zero term in the linear expansion for some vector in Group 3; otherwise it belongs to Group 1.

We observe that after the above process, there are $K_1$ Type-1 columns, $K_2$ Type-2 columns, and at most $K_2$ Type-3 columns. This is because when we label each Type-3 column, we always set *inactive* at least 1 vector in Group 2.

To deal with the case that $\boldsymbol{A}$'s rank $K'$ is less than $K$, we arbitrarily label $K - K'$ unlabeled column vectors in Group 3 as Type-2 columns to make $K_1 + K_2 = K$; we can also arbitrarily mark another $K_2 - K_3$ unlabeled column vectors in Group 3 as Type-3 columns to make $K_2 = K_3$. It is easy to see that after this padding, the selected Type-1, Type-2, and Type-3 columns satisfy the desired properties.

Given the fact that $\mathcal{S} = \mathcal{A}$, we focus on $\mathcal{S}$ and prove the following two lemmas.

**Lemma 4.** *There are in total no more than* $\sum_{K_1+K_2=K} \binom{m}{K_1}\binom{m-K_1}{K_2}\binom{m-K_1-K_2}{K_2} \leq 2m^{2K}$ *coding structures corresponding to all* $K \times m$ *coding matrices.*

*Proof.* We can see that we have at most $\binom{m}{K_1}$ ways to choose the $K_1$ Type-1 columns, $\binom{m-K_1}{K_2}$ ways to choose the $K_2$ Type-2 columns among the remaining $m - K_1$ columns, and $\binom{m-K_1-K_2}{K_2}$ ways to choose the $K_3 = K_2$ Type-3 columns among the remaining $m - K_1 - K_2$ columns. Hence, the total number of coding structures is no more than

$$\begin{aligned} \sum_{K_2=0}^{K} m^K (m-K)^{K_2} &\leq \sum_{K_2=0}^{K-1} m^K (m-K)^{K_2} + m^{2K} \\ &\leq \frac{m^K((m-K)^{K+1}-1)}{m-K-1} + m^{2K} \leq 2m^{2K}. \end{aligned} \tag{11}$$

□

To determine whether a given matrix does not satisfy a problem instance, it is sufficient for us to simply look at the coding structure this matrix belongs to. We say that a coding structure satisfies a client or a problem instance if and only if there exists some matrix belonging to this coding structure that satisfies the client or the problem instance.

**Lemma 5.** *The probability that all $n$ clients are satisfied by a coding structure $S(J^{(1)}, J^{(2)}, J^{(3)})$ can be upper bounded by*

$$\Pr\{S(J^{(1)}, J^{(2)}, J^{(3)}) \text{ can satisfy all } n \text{ clients}\} \leq \begin{cases} [1-p^{2K}]^n, & p \leq \frac{\sqrt{5}-1}{2}, \\ [1-(1-p)^K]^n, & p > \frac{\sqrt{5}-1}{2}. \end{cases} \tag{12}$$

*Proof.* We denote the coding structure $S(J^{(1)}, J^{(2)}, J^{(3)})$ by $S$ for short. We first notice that if a client $i$ has the following connection pattern: $j' \notin R_i$ for any column $j' \in J^{(1)}$ and $j'' \in R_i$ for any column $j'' \in J^{(2)} \cup J^{(3)}$, then client $i$ cannot be satisfied by any matrix in the coding structure $S$. Indeed, if a client $i$ has the above connection pattern, then clearly:

• Client $i$ has all messages indexed by $J^{(1)}$ as side information and cannot be satisfied by messages in $J^{(1)}$.

• Client $i$ cannot decode any message in $J^{(2)} \cup J^{(3)}$ according to the decoding criterion, since any column vector indexed by $J^{(2)} \cup J^{(3)}$ is in the linear space spanned by all other vectors in $J^{(2)} \cup J^{(3)}$ from the definitions of $J^{(2)}$ and $J^{(3)}$.

• Client $i$ cannot decode a message not indexed by $J^{(1)}$, $J^{(2)}$, and $J^{(3)}$, because column vectors indexed by $J^{(2)}$ contains a basis for the submatrix that is obtained from removing columns of $J^{(1)}$, and this implies that the messages not indexed by $J^{(1)}$, $J^{(2)}$, and $J^{(3)}$ are in the space spanned by vectors indexed by $J^{(2)}$.

Next, we can lower bound the probability that client $i$ is not satisfied by $S$ by calculating the probability that event $\{j' \notin R_i, \forall j' \in J^{(1)}, \text{ and } j'' \in R_i, \forall j'' \in J^{(2)} \cup J^{(3)}\}$ happens.

$$\begin{aligned} &\Pr\{\text{client } i \text{ is not satisfied by } S\} \\ &\geq \Pr\{j' \notin R_i, \forall j' \in J^{(1)}, \text{ and } j'' \in R_i, \forall j'' \in J^{(2)} \cup J^{(3)}\} \\ &\geq (1-p)^{K_1} p^{2K_2}. \end{aligned} \tag{13}$$

Therefore, we can upper bound the probability that all $n$ clients are satisfied by structure $S$ as follows.

$$\Pr\{\text{all } n \text{ clients are satisfied by } S\} \leq [1-(1-p)^{K_1} p^{2K_2}]^n. \tag{14}$$

Note that for $p \leq (\sqrt{5}-1)/2$ we have $1-p \geq p^2$, and for $p > (\sqrt{5}-1)/2$ we have $1-p < p^2$. So that the result follows from eq. (14) and the fact that $K_1 + K_2 = K$. $\square$

A lower bound is shown in the following theorem.

**Theorem 5.** *For pliable index coding over random graph $B(m,n,p)$ ($m = \mathcal{O}(n^{\delta})$ for some constant $\delta$), with probability at least $1 - \mathcal{O}(1/n^2)$, the linear pliable index code length can be lower bounded as follows:*

$$K \geq \begin{cases} \frac{\log(n)}{4\log(1/p)}, & p \leq \frac{\sqrt{5}-1}{2}, \\ \frac{\log(n)}{2\log[1/(1-p)]}, & p > \frac{\sqrt{5}-1}{2}. \end{cases} \tag{15}$$

*Proof.* According to Lemmas 4 and 5, we can see that the probability a random graph $B(m,n,p)$ can be satisfied by a pliable index code of length $K = c(p)\log(n)$ (the parameter $c(p) = \frac{1}{4\log(1/p)}$ for $p \leq \frac{\sqrt{5}-1}{2}$ and $c(p) = \frac{1}{2\log[1/(1-p)]}$ for $p > \frac{\sqrt{5}-1}{2}$) is at most

$$\begin{aligned} &2m^{2K}[1 - \frac{1}{2^{\frac{K}{2c(p)}}}]^n = 2m^{2c(p)\log(n)}[1 - \frac{1}{2^{\frac{\log(n)}{2}}}]^n \\ &\leq [2^{2c(p)\log(n)\log(m)+1}]/e^{\sqrt{n}} \leq \mathcal{O}(1/n^2). \end{aligned} \tag{16}$$

$\square$

From Theorem 5, we distinguish the following special cases for the lower bound depending on $p$.

• $p \leq \mathcal{O}(1/n^{\alpha})$ or $1 - p \leq \mathcal{O}(1/n^{\alpha})$ for some constant $\alpha$: this is the sparse or dense case and we get $\Omega(1)$ lower bound from Theorem 5. To explain this, consider two extreme cases. The first one is a sparse case where each client requires exactly one different message and has all others as side information. Thus, we only need to transmit a linear combination of all the messages, such that each client can decode her required message. The other one is a dense case where each client has only one side-information message and requires any new one from the remaining messages. In this case, we can use 2 arbitrary uncoded transmissions to satisfy all clients.

• Constant $p$: in this case we achieve $K \geq \Omega(\log(n))$ from Theorem 5, namely, the random instance $B(m,n,p)$ almost surely needs linear code length of $\Omega(\log(n))$. In particular, when $p = (\sqrt{5}-1)/2 \approx 0.618$, *the Golden Ratio*, this lower bound achieves maximum $0.36\log(n)$ among all $p$.

### B. Upper Bound

To prove an upper bound, we propose a simple coding scheme that achieves code length of $\mathcal{O}(\log(n))$ with high probability.

Given a constant $p$ and $m = \mathcal{O}(n^\delta)$ for some constant $\delta$, we construct the coding matrix $\boldsymbol{A}$ as follows:

$$\boldsymbol{A} = \begin{bmatrix} 1 & 1 & \cdots & 1 & 0 & 0 & \cdots & 0 & \cdots & 0 & 0 & \cdots & 0 & 0 & 0 & \cdots & 0 \\ 0 & 0 & \cdots & 0 & 1 & 1 & \cdots & 1 & \cdots & 0 & 0 & \cdots & 0 & 0 & 0 & \cdots & 0 \\ \vdots & \vdots & \vdots & \vdots & \vdots & \vdots & \vdots & \vdots & \ddots & \vdots & \vdots & \vdots & \vdots & \vdots & \vdots & \vdots & \vdots \\ 0 & 0 & \cdots & 0 & 0 & 0 & \cdots & 0 & \cdots & 1 & 1 & \cdots & 1 & 0 & 0 & \cdots & 0 \end{bmatrix}. \tag{17}$$

The matrix has $\frac{3}{\log(e/(e-1))}\log(n)$ rows. In each row, we have a constant weight: $1/p$ 1s and 0s for other elements[4]. In any two rows, the 1s are non-overlapping. The probability that a client $i$ is satisfied by the first row can be upper bounded by the following equation.

$$\Pr\{\text{client } i \text{ is satisfied by the first row}\} = \binom{1/p}{1} p(1-p)^{1/p-1} \geq 1/e. \tag{18}$$

Note that since 1s in any two rows of the coding matrix do not overlap, we can calculate the probability that a client $i$ is satisfied by the coding matrix $\boldsymbol{A}$ as:

$$\Pr\{\text{client } i \text{ is satisfied by the coding matrix } \boldsymbol{A}\} \geq 1-(1-1/e)^{\frac{3}{\log(e/(e-1))}\log(n)} \geq 1-1/n^3. \tag{19}$$

Hence, the probability that all clients are satisfied can be bounded as:

$$\Pr\{\text{all clients are satisfied by the coding matrix } \boldsymbol{A}\} \geq (1-\frac{1}{n^3})^n \geq 1-\frac{1}{n^2}. \tag{20}$$

Therefore, we have the following result.

**Theorem 6.** *For pliable index coding over random graph $B(m,n,p)$ ($m = \mathcal{O}(n^\delta)$ for some constant $\delta$) with constant $p$, we can achieve the optimal linear pliable index code length $K = \Theta(\log(n))$ almost surely.*

To illustrate how the lower and upper bounds change with the probability $p$, we plot the relationship between them in Fig. 3.

[4]We simply treat $1/p$ as integers, which does not change the problem essentially.

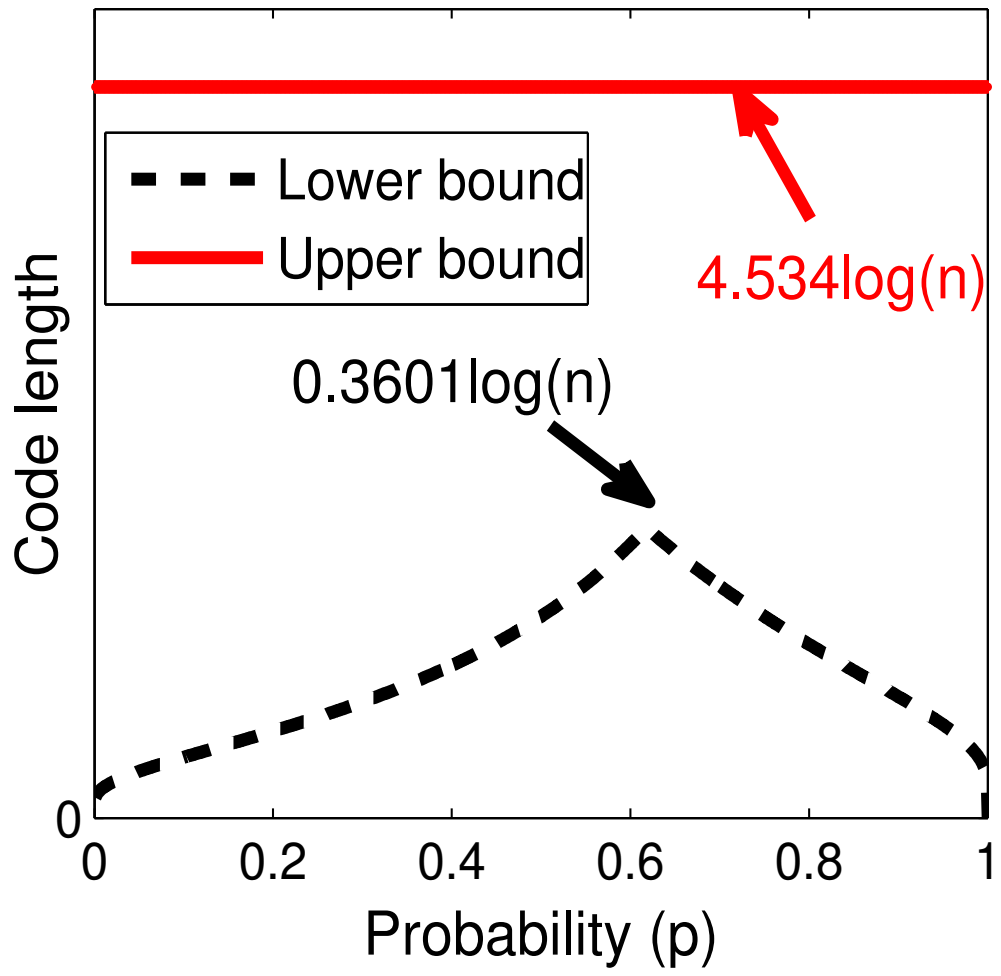

Fig. 3: Lower and upper bounds of pliable index coding over random graphs.

# VIII. Discussion

In this section, we make two observations: one on the field size for the optimal solution and the other on a connection with the minrank problem.

## A. Field Size

We show through an example that a binary code is not sufficient to achieve the optimal code length. Consider the following instance with $m = 4$ and $n = 10$:

- $R_1 = \{1\}, R_2 = \{2\}, R_3 = \{3\}, R_4 = \{4\}, R_5 = \{1,2\}, R_6 = \{1,3\}, R_7 = \{1,4\}, R_8 = \{2,3\}, R_9 = \{2,4\}, R_{10} = \{3,4\}$.

This instance contains clients with request sets of all 1-message and 2-message subsets. We can easily see that the optimal code length is 2, e.g., $b_1 + b_2 + b_4$ and $b_2 + b_3 + 2b_4$, in $\mathbf{F}_3$. However, we cannot find a binary code of length 2, because we have all 1-message and 2-message request sets, requiring $\boldsymbol{a}_j \neq (0,0)^T$, for $j = 1,2,3,4$ and $\boldsymbol{a}_j \neq \boldsymbol{a}_{j'}$, for $j \neq j'$. But, we have only 3 non-zero vectors $(1,0)^T, (0,1)^T, (1,1)^T$. It is not possible to assign these 3 non-zero vectors to 4 columns so as to satisfy all clients.

This example extends to that at least a field size $m-1$ of coding coefficients is needed to achieve the optimal code length for all instances with $m$ messages. We consider an instance with $m$ messages and $n = m + \binom{m}{2}$ clients, where the clients have all 1-message and 2-message request sets. Namely, the clients' request sets are $\{j\}$ and $\{j_1, j_2\}$, for any $j \in [m]$ and $j_1, j_2 \in [m]$.

Assume we use finite field $\boldsymbol{F}_q$ to realize coding. According to our decoding criterion, we need every coding vector to be nonzero and any pair of the coding vectors to be linearly independent.

• If the coding vector contains 0, then there will be 2 of them: $(1,0)^T$ and $(0,1)^T$ since any other vector in the form of $(x,0)^T$ ($x \in \boldsymbol{F}_q$) is linearly dependent with $(1,0)^T$ and similarly, $(0,x)^T$ ($x \in \boldsymbol{F}_q$) is linearly dependent with $(0,1)^T$.

• If the coding vector is in the form $(x,y)^T$, $x,y \in \boldsymbol{F}_q, x,y \neq 0$, then there are in total $(q-1)^2$ such vectors. However, $(x,y)^T$ is linearly dependent with $z(x,y)^T$, for $z \in \boldsymbol{F}_q$. There are in total $(q-1)$ distinct $z(x,y)^T$ vectors, so the total number of pair-wise independent vectors is $(q-1)^2/(q-1) = (q-1)$.

Therefore, we need $2 + (q-1) \geq m$ in order to satisfy these clients, resulting in $q \geq m-1$.

### B. Minrank

In index coding, the optimal linear code length is shown to equal to the minrank, which is the minimum rank of a mixed matrix (some of whose elements are to be determined) associated with the side-information graph [11]. In a similar way, we can characterize the pliable index coding problem using the minimum rank of a mixed matrix associated with the bipartite graph.

We say that a matrix $\boldsymbol{G} \in \mathbf{F}_q^{n\times m}$ fits the pliable index coding problem $(m, n, \{R_i\}_{i\in[n]})$ if in the $i$-th row ($\forall i \in [n]$):

- among all $j \in R_i$, there exists one and only one $j^* \in R_i$, such that $g_{ij^*} = 1$, and other $g_{ij} = 0$ for any $j \in R_i \backslash \{j^*\}$;
- for $j \in S_i$, $g_{ij}$ can be any element in $\mathbf{F}_q$.

Let us denote by $\mathcal{G}$ the set of all matrices fitting the pliable index coding problem $(m, n, \{R_i\}_{i\in[n]})$, and by minrank($\mathcal{G}$) the minimum rank among all the matrices $\boldsymbol{G} \in \mathcal{G}$. In other words, minrank$(\mathcal{G}) = \min_{\boldsymbol{G}\in\mathcal{G}} \text{rank}(\boldsymbol{G})$, where rank($\boldsymbol{G}$) denotes the rank of matrix $\boldsymbol{G}$. The following theorem characterizes the optimal coding length:

**Theorem 7.** *The optimal linear code length of the pliable index coding instance* $(m, n, \{R_i\}_{i\in[n]})$ *equals to minrank*$(\mathcal{G})$.

*Proof.* First, let us prove that a linear code with length $K = \text{minrank}(\mathcal{G})$ exists. Assume that a matrix $\boldsymbol{G} \in \mathcal{G}$ achieves rank $K$. Without loss of generality, let us also assume that the first $K$ rows of $\boldsymbol{G}$ are linearly independent. For the encoding process, we define the coding matrix $\boldsymbol{A}$ to be the first $K$ rows of $\boldsymbol{G}$. For matrix $\boldsymbol{G}$, there is one and only one $j^* \in R_i$, such that $g_{ij^*} = 1$, and other $g_{ij} = 0$ for $j \in R_i \backslash \{j^*\}$; so that column $\boldsymbol{g}_{j^*}$ cannot be expressed as a linear combination of $\{\boldsymbol{g}_j\}_{j\in R_i \backslash \{j^*\}}$. Since all the rows of $\boldsymbol{G}$ are linear combinations of the first $K$ rows, column $\boldsymbol{a}_{j^*}$ cannot be expressed as a linear combination of $\{\boldsymbol{a}_j\}_{j \in R_i \setminus \{j^*\}}$ either. As a result, the decoding criterion holds for client $i$ and message $j^*$ can be decoded by client $i$.

Next, let us prove that for any linear code with a $K \times m$ coding matrix $\boldsymbol{A}$ in filed $\mathbf{F}_q$ has a code length $K \geq \text{minrank}(\mathcal{G})$. We show that using the coding matrix $\boldsymbol{A}$, we can build a matrix $\boldsymbol{G} \in \mathbf{F}_q^{n \times m}$ with rank at most $K$ that fits the index coding problem. To show this, we use the following claim.

**Claim 1**: If for client $i$, the message $j^*$ can be decoded, then the row vector $e_{j^*}^T$ is in the span of $\{\boldsymbol{\alpha}_l^T : l \in [K]\} \cup \{e_j^T : j \in S_i\}$, where $e_j^T$ is a row vector with all 0s, except a 1 in the $j$-th position and $\boldsymbol{\alpha}_l^T$ represents the $l$-th row of matrix $\boldsymbol{A}$.

This claim shows that $e_{j^*}^T$ is in the span of the union of row space of $\boldsymbol{A}$ and the side-information space. The proof of this claim can be found in [11].

For each client $i$, the claim states that $e_{j^*}^T = \sum_{l=1}^{K} \lambda_l \boldsymbol{\alpha}_l^T + \sum_{j \in S_i} \mu_j e_j^T$ for some $\lambda_l, \mu_j$ in field $\mathbf{F}_q$. To construct $\boldsymbol{G}$, we define the $i$-th row of $\boldsymbol{G}$, $\gamma_i^T$, to be the linear combination $\sum_{l=1}^{K} \lambda_l \boldsymbol{\alpha}_l^T$. Or equivalently, we have $\gamma_i^T = \sum_{l=1}^{K} \lambda_l \boldsymbol{\alpha}_l^T = e_{j^*}^T - \sum_{j \in S_i} \mu_j e_j^T$. This shows that $\gamma_i^T$ has value 1 at position $j^*$, $-\mu_j$ at position $j \in S_i$, and 0 at positions indexed by $R_i \setminus \{j^*\}$.

Therefore, we have shown that $K \geq \text{rank}(\boldsymbol{G}) \geq \text{minrank}(\mathcal{G})$. □

### *C. Vector Pliable Index Coding*

Up to now in this paper, the problem definition and the subsequent analysis are based on scalar pliable index coding, where we consider each message to be an element in a finite field $\mathbf{F}_q$. An interesting consideration is the extension to the vector pliable index coding where a message is considered to be an element in the field $\mathbf{F}_{q^L}$, or a vector of length $L$ in field $\mathbf{F}_q$. In this case, the $m$ messages are represented as $\boldsymbol{b}_1 = (b_{1,1}, b_{1,2}, \ldots, b_{1,L})$, $\boldsymbol{b}_2 = (b_{2,1}, b_{2,2}, \ldots, b_{2,L})$, ..., $\boldsymbol{b}_m = (b_{m,1}, b_{m,2}, \ldots, b_{m,L})$. For easy of exposition, we call each element of a message vector in the finite field $\mathbf{F}_q$ a sub-message, i.e., $b_{j,l}$ for all $j \in [m], l \in [L]$. The side information set and the request set are the same for client $i$ as in the scalar pliable index coding case, i.e., $S_i \subseteq [m]$ and $R_i = [m] \setminus S_i$. A client $i$ is satisfied as long as she decodes a message $\boldsymbol{b}_j = (b_{j,1}, b_{j,2}, \ldots, b_{j,L})$ for some $j \in R_i$ or $t$ unknown messages $\boldsymbol{b}_{j_1}, \boldsymbol{b}_{j_2}, \ldots, \boldsymbol{b}_{j_t}$ for some $j_1, j_2, \ldots, j_t \in R_i$ for the $t$-requests case. The goal is to minimize the number of transmissions, where each transmission is a vector of length $L$ with elements in field $\mathbf{F}_q$.

The linear encoding process can be represented by the equation $\boldsymbol{A}^{\ddagger}\boldsymbol{b}^{\ddagger} = \boldsymbol{x}^{\ddagger}$, where the coding matrix $\boldsymbol{A}^{\ddagger}$ of size $K^{\ddagger} \times mL$ is defined as

$$\begin{bmatrix} a_{1,1,1} & a_{1,1,2} & \cdots & a_{1,1,L} & a_{1,2,1} & a_{1,2,2} & \cdots & a_{1,2,L} & \cdots & a_{1,m,1} & a_{1,m,2} & \cdots & a_{1,m,L} \\ a_{2,1,1} & a_{2,1,2} & \cdots & a_{2,1,L} & a_{2,2,1} & a_{2,2,2} & \cdots & a_{2,2,L} & \cdots & a_{2,m,1} & a_{2,m,2} & \cdots & a_{2,m,L} \\ \vdots & \vdots & \vdots & \vdots & \vdots & \vdots & \vdots & \vdots & \ddots & \vdots & \vdots & \vdots & \vdots \\ a_{K^{\ddagger},1,1} & a_{K^{\ddagger},1,2} & \cdots & a_{K^{\ddagger},1,L} & a_{K^{\ddagger},2,1} & a_{K^{\ddagger},2,2} & \cdots & a_{K^{\ddagger},2,L} & \cdots & a_{K^{\ddagger},m,1} & a_{K^{\ddagger},m,2} & \cdots & a_{K^{\ddagger},m,L} \end{bmatrix};$$

the vector $\boldsymbol{b}^{\ddagger}$ is obtained by concatenating all sub-messages in $\boldsymbol{b}_j$ for all $j \in [m]$: $\boldsymbol{b}^{\ddagger} = (b_{1,1}, b_{1,2}, \dots, b_{1,L}, b_{2,1}, b_{2,2}, \dots, b_{2,L}, \dots, b_{m,1}, b_{m,2}, \dots, b_{m,L})^T$; and $\boldsymbol{x}^{\ddagger}$ is the vector of $K^{\ddagger}$ transmitted sub-messages. The coefficient $a_{k,j,l}$ with three subscripts, an element of $\boldsymbol{A}^{\ddagger}$, correspond to the row $k$ of the matrix, the message index $j$, and the $l$-th sub-message in the message vector $\boldsymbol{b}_j$. We will also write the matrix $\boldsymbol{A}^{\ddagger}$ in its column vector form:

$$\boldsymbol{A}^{\ddagger} = (\boldsymbol{a}_{1,1}, \boldsymbol{a}_{1,2}, \dots, \boldsymbol{a}_{1,L}, \boldsymbol{a}_{2,1}, \boldsymbol{a}_{2,2}, \dots, \boldsymbol{a}_{2,L}, \dots, \boldsymbol{a}_{m,1}, \boldsymbol{a}_{m,2}, \dots, \boldsymbol{a}_{m,L}), \tag{21}$$

where $\boldsymbol{a}_{j,l}$ denotes the column of $\boldsymbol{A}^{\ddagger}$ corresponding to the sub-message $b_{j,l}$. Note that in this encoding process, we can encode sub-messages between message vectors or inside a message vector. When we express the encoding process in this matrix form, elements of $\boldsymbol{A}^{\ddagger}$, $\boldsymbol{b}^{\ddagger}$, and $\boldsymbol{x}^{\ddagger}$ are in the finite field $\mathbf{F}_q$. Using this transformation, we can use many properties of the scalar pliable index coding. We then define the (equivalent) code length or the (equivalent) number of transmissions as $K = K^{\ddagger}/L$ to eliminate the effect of the vector length.

Similarly to the scalar pliable index coding, given $\boldsymbol{A}^{\ddagger}$, $\boldsymbol{x}^{\ddagger}$, and $\{\boldsymbol{b}_j | j \in S_i\}$, the decoding process for client $i$ is to solve $\boldsymbol{A}^{\ddagger}\boldsymbol{b}^{\ddagger} = \boldsymbol{x}^{\ddagger}$ to get a unique solution of $\boldsymbol{b}_j$ for some $j \in R_i$, or unique solutions $\boldsymbol{b}_{j_1}, \boldsymbol{b}_{j_2}, \dots, \boldsymbol{b}_{j_t}$ for some $j_1, j_2, \dots, j_t \in R_i$ for the $t$-requests case. Clearly, client $i$ can remove her side information messages, i.e., can create $x_k^{\ddagger(i)} = x_k^{\ddagger} - \sum_{j\in S_i, l\in[L]} a_{kjl} b_{jl}$ from all $k \in [K^{\ddagger}]$. As a result, client $i$ needs to solve the equations

$$\boldsymbol{A}^{\ddagger}_{R_i} \boldsymbol{b}^{\ddagger}_{R_i} = \boldsymbol{x}^{\ddagger(i)}, \tag{22}$$

to retrieve any one message (or $t$ messages) she does not have, where $\boldsymbol{A}^{\ddagger}_{R_i}$ is the sub-matrix of $\boldsymbol{A}^{\ddagger}$ with columns corresponding to messages in $R_i$; $\boldsymbol{b}^{\ddagger}_{R_i}$ is the message vector with elements corresponding to messages in $R_i$; and $\boldsymbol{x}^{\ddagger(i)}$ is a $K^{\ddagger}$-dimensional column vector with element $x_k^{\ddagger(i)}$.

Note that if we treat each sub-message separately, then the decoding criterion still holds for a single sub-message, i.e., a client $i$ can decode a sub-message $b_{j,l}$ for some $j \in R_i, l \in [L]$ if and only if the column $\boldsymbol{a}_{j,l}$ corresponding to $b_{j,l}$ is not in the space spanned by columns of $\boldsymbol{A}^{\ddagger}_{R_i}$ other than $\boldsymbol{a}_{j,l}$. However, in the vector pliable index coding, a client $i$ needs to be able to decode all sub-messages for some message vector $\boldsymbol{b}_j$, $j \in R_i$, namely, for any $b_{j,l}$, $l \in [L]$ to satisfy the decoding criterion. In this case, we can derive an equivalent criterion based on the decoding criterion in Lemma 1.

**Lemma 6.** *A client $i$ can decode a message $\boldsymbol{b}_j$, $j \in R_i$, if and only if the following two conditions are satisfied:*
*1) the space spanned by columns of $\boldsymbol{A}^{\ddagger}_{R_i}$ corresponding to sub-messages $b_{j,l}$, $\forall l \in [L]$, span{ $\boldsymbol{a}_{j,1}, \boldsymbol{a}_{j,2}, \ldots, \boldsymbol{a}_{j,L}$}, has dimension $L$; in other words, columns $\boldsymbol{a}_{j,1}, \boldsymbol{a}_{j,2}, \ldots, \boldsymbol{a}_{j,L}$ are linearly independent and form a basis of the space;*
*2) any non-zero vector in the space spanned by columns of $\boldsymbol{A}^{\ddagger}_{R_i}$ corresponding to sub-messages $b_{j,l}$, $\forall l \in [L]$ is not in the space spanned by columns of $\boldsymbol{A}^{\ddagger}_{R_i}$ other than those corresponding to sub-messages $b_{j,l}$, $\forall l \in [L]$:*

$$\boldsymbol{v} \notin \textit{span}\{\boldsymbol{a}_{j',l'} | j' \in R_i \backslash \{j\}, l' \in [L]\}, \forall \boldsymbol{v} \in \textit{span}\{\boldsymbol{a}_{j,1}, \boldsymbol{a}_{j,2}, \ldots, \boldsymbol{a}_{j,L}\}, \boldsymbol{v} \neq 0. \tag{23}$$

*Proof.* We first prove the necessary condition. Assume that a client $i$ can decode any sub-message $b_{j,l}$ for all $l \in [L]$. Hence, according to the decoding criterion in Lemma 1, $\boldsymbol{a}_{j,l}$ is not in the space spanned by $\boldsymbol{a}_{j,1}, \ldots, \boldsymbol{a}_{j,l-1}, \boldsymbol{a}_{j,l+1}, \ldots, \boldsymbol{a}_{j,L}$. Note that this holds for all $l \in [L]$, then the condition 1) holds.

For condition 2), assume that some non-zero vector $\boldsymbol{v} = \gamma_1 \boldsymbol{a}_{j,1} + \gamma_2 \boldsymbol{a}_{j,2} + \ldots + \gamma_L \boldsymbol{a}_{j,L}$ ($\gamma_l$ for all $l \in [L]$ are coefficients in the field $\mathbf{F}_q$) is in the space spanned by $\{\boldsymbol{a}_{j',l'} | j' \in R_i \backslash \{j\}, l' \in [L]\}$. Without loss of generality, let us denote by $\boldsymbol{v}_1, \boldsymbol{v}_2, \ldots, \boldsymbol{v}_{L^c}$ a basis of the space $\{\boldsymbol{a}_{j',l'} | j' \in R_i \backslash \{j\}, l' \in [L]\}$ with dimension $L^c$. Therefore, the vector $\boldsymbol{v}$ can be expressed as a linear combination of the basis:

$$\boldsymbol{v} = \gamma_1 \boldsymbol{a}_{j,1} + \gamma_2 \boldsymbol{a}_{j,2} + \ldots + \gamma_L \boldsymbol{a}_{j,L} = \beta_1 \boldsymbol{v}_1 + \beta_2 \boldsymbol{v}_2 + \ldots + \beta_{L^c} \boldsymbol{v}_{L^c}, \tag{24}$$

where $\beta_l$ for all $l \in [L^c]$ are coefficients in the field $\mathbf{F}_q$. Since $\boldsymbol{v}$ is non-zero, some $\gamma_{l^*}$ is non-zero, and hence, the column $\boldsymbol{a}_{j,l^*}$ is in the space spanned by other columns of $\boldsymbol{A}^{\ddagger}_{R_i}$, which results in a contradiction for the client $i$ to decode $b_{j,l^*}$ according to the decoding criterion in Lemma 1.

We next prove the sufficient condition. Assume that the two conditions hold and some sub-message $b_{j,l}$ for some $l \in [L]$ cannot be decoded. Using the decoding criterion in Lemma 1, we can express $\boldsymbol{a}_{j,l}$ as a linear combination of $\boldsymbol{a}_{j,1}, \ldots, \boldsymbol{a}_{j,l-1}, \boldsymbol{a}_{j,l+1}, \ldots, \boldsymbol{a}_{j,L}$ and the basis of the space $\{\boldsymbol{a}_{j',l'} | j' \in R_i \backslash \{j\}, l' \in [L]\}$ with dimension $L^c$, denoted by $\boldsymbol{v}_1, \boldsymbol{v}_2, \ldots, \boldsymbol{v}_{L^c}$:

$$\boldsymbol{a}_{j,l} = \gamma_1 \boldsymbol{a}_{j,1} + \ldots + \gamma_{l-1} \boldsymbol{a}_{j,l-1} + \gamma_{l+1} \boldsymbol{a}_{j,l+1} + \ldots + \gamma_L \boldsymbol{a}_{j,L} + \beta_1 \boldsymbol{v}_1 + \beta_2 \boldsymbol{v}_2 + \ldots + \beta_{L^c} \boldsymbol{v}_{L^c}, \quad (25)$$

where $\gamma_{l'}$ for all $l' \in [L] \backslash \{l\}$ and $\beta_{l''}$ for all $l'' \in [L^c]$ are coefficients in the field $\mathbf{F}_q$. If all $\beta_{l''}$ are 0, then $\boldsymbol{a}_{j,l}$ is expressed as a linear combination of $\boldsymbol{a}_{j,1}, \ldots, \boldsymbol{a}_{j,l-1}, \boldsymbol{a}_{j,l+1}, \ldots, \boldsymbol{a}_{j,L}$, resulting in a contradiction to condition 1). Therefore, $\beta_{l''}$ cannot be all 0s. In this case, the vector $\boldsymbol{a}_{j,l} - \gamma_1 \boldsymbol{a}_{j,1} - \ldots - \gamma_{l-1} \boldsymbol{a}_{j,l-1} - \gamma_{l+1} \boldsymbol{a}_{j,l+1} - \ldots - \gamma_L \boldsymbol{a}_{j,L}$ is expressed as $\beta_1 \boldsymbol{v}_1 + \beta_2 \boldsymbol{v}_2 + \ldots + \beta_{L^c} \boldsymbol{v}_{L^c}$, which contradicts condition 2). The lemma is proved. □

This lemma extends the relationship between column vectors in scalar pliable index coding to the relationship between linear subspaces. If we consider each original message vector corresponds to a linear subspace in the coding matrix, then this lemma states that the decodable message $j$ corresponds to a $L$-dimensional linear subspace and this subspace is independent of the sum of the other subspaces corresponding to the request set other than $j$.

Any upper bound of the optimal number of transmissions for scaler pliable index coding is also an upper bound of the optimal number of equivalent transmissions for the vector pliable index coding. This can be done by treating each sub-message of a message vector sequentially and encoding only the sub-messages at the same position $l \in [L]$ in the message vectors. This means that vector pliable index coding cannot do worst than scalar pliable index coding.

An interesting question is the lower bounds. Based on the proposed criterion in Lemma 6, we show that the lower bounds achieved using scalar pliable index coding are also the lower bounds for the vector pliable index coding, in both the worst case and the average case.

**Corollary 1.** *In a complete instance $(m, n, \{R_i\}_{i \in [n]})$ for vector pliable index coding with vector length $L$, the optimal number of equivalent transmissions is $\Omega(\log(n))$.*

**Corollary 2.** *In a complete instance $(m, n, \{R_i\}_{i \in [n]}, t)$ for vector pliable index coding with vector length $L$, the optimal number of equivalent transmissions is $\Omega(t + \log(n))$.*

**Corollary 3.** *For vector pliable index coding with vector length $L$ over random graph $B(m, n, p)$ ($m = \mathcal{O}(n^{\delta})$ for some constant $\delta$), with probability at least $1 - \mathcal{O}(1/n^2)$, the equivalent linear* *pliable index code length can be lower bounded as follows:*

$$K \geq \begin{cases} \frac{\log(n)}{4\log(1/p)}, & p \leq \frac{\sqrt{5}-1}{2}, \\ \frac{\log(n)}{2\log[1/(1-p)]}, & p > \frac{\sqrt{5}-1}{2}. \end{cases} \tag{26}$$

Proofs of these corollaries are similar to the proofs for the scalar pliable index coding and are provided in Appendix B.

## IX. Related Work

The index coding problem was first introduced in [3], [11] and shown to be NP-hard [10], [11], [12]. The optimal linear code length is shown as minimum rank of a family of matrices and the optimal linear code length has a sandwich property, namely, the optimal code length is lower bounded and upper bounded by clique number and chromatic number of some specifically defined graphs [11]. Various techniques, e.g., linear programming [5], interference alignment [15], information theory [16], network coding and matroid theory [17], have also been used to analyze the index coding problem. In [4], the insufficiency of linear codes to achieve the optimum is shown by some special examples. The equivalence of index coding and network coding is studied in [17], [18]. In [16], [12], the capacity and rate of index coding are studied through information theoretical analysis. In addition, several aspects of index coding problem are also investigated in the literature, such as the complementary index coding problem [19], security of index coding [20], efficient algorithms [6], and index coding with outerplanar side information [21].

The analysis of index coding over random graphs characterizes "typical" or "average" performance of index coding problem. One can refer to [14] to get more details about random graphs. The work in [22] shows that minimum length of the scalar index code for a random graph is almost surely $\Omega(\sqrt{n})$. A recent work improves this bound for scalar index coding by showing that the minrank achieves $\Theta(n/\log(n))$ almost surely [13].

Pliable index coding was introduced in [7], [8]. The work in [7], [8] has shown that, although in the worst case, for conventional index coding, we may require $\Omega(n)$ transmissions, for pliable index coding we require at most $\mathcal{O}(\log^2(n))$, i.e., we can be exponentially more efficient. However, to achieve the optimal coding length is still NP-hard [7]. For pliable index coding with $t$-requests, the work in [8] has shown that an upper bound is $\mathcal{O}(t\log(n)+\log^3(n))$ for the code length. This result is derived from probabilistic arguments and an appropriate utilization of MDS codes. Compared with previous work on pliable index coding where performance guarantee is achieved only using random algorithms and on average sense [7], [8], our work in this paper is to design a deterministic polynomial-time algorithm.

## X. Numerical Results

In this section, we conduct numerical experiments on our proposed algorithms. We first evaluate the performance of our BinGreedy algorithm by comparing with the algorithm in [7], and then evaluate the optimality gap with respect to the minrank solution in Section VIII. We finally evaluate the BinGreedyT algorithm performance for $t$-requests case.

### A. Performance Comparison

We compare the performance of our proposed algorithm BinGreedy with *RANDCOV*, which is a randomized algorithm proposed in [7]. RANDCOV is the current state-of-the art alternative and was theoretically shown to achieve an *average performance* upper bounded by $\mathcal{O}(\log^2(n))$ with respect to the random code realization.

In our simulations, we set the number of messages $m$ to be $n^{0.75}$ and numerically investigate how the code length changes with the number of clients $n$. We randomly generate $100$ pliable index coding bipartite graph instances for each $n$, by connecting each client and each message with probability $0.3$ in the homogeneous case; and connecting equal number of clients with each message with probabilities $0.05, 0.15, \ldots, 0.95$ in the heterogeneous case.

Fig. 4 shows the code length varying with $n$ (note that the horizontal axis is in logarithmic scale). We can see that on average (averaged over $100$ instances for the same $n$) and in the worst case, the proposed BinGreedy algorithm outperforms RANDCOV by $20\%$-$35\%$ in terms of the code length for homogeneous instances. In heterogeneous instances, the proposed BinGreedy algorithm outperforms the existing randomized algorithm by $20\%$-$50\%$. We also observe that for heterogeneous instances, we need more transmissions than the homogeneous instances of the same size. As seen in the figure, for homogeneous instances, the code length increases almost linearly with $\log(n)$; while for heterogeneous instances, the code length increases super linearly with $\log(n)$.

In contrast to the randomness of RANDCOV, our proposed BinGreedy algorithm runs deterministically and we expect more robustness. Indeed, we can see from Fig. 4 that the difference between best case and worst case instances is much larger for RANDCOV than that for the proposed BinGreedy algorithm.

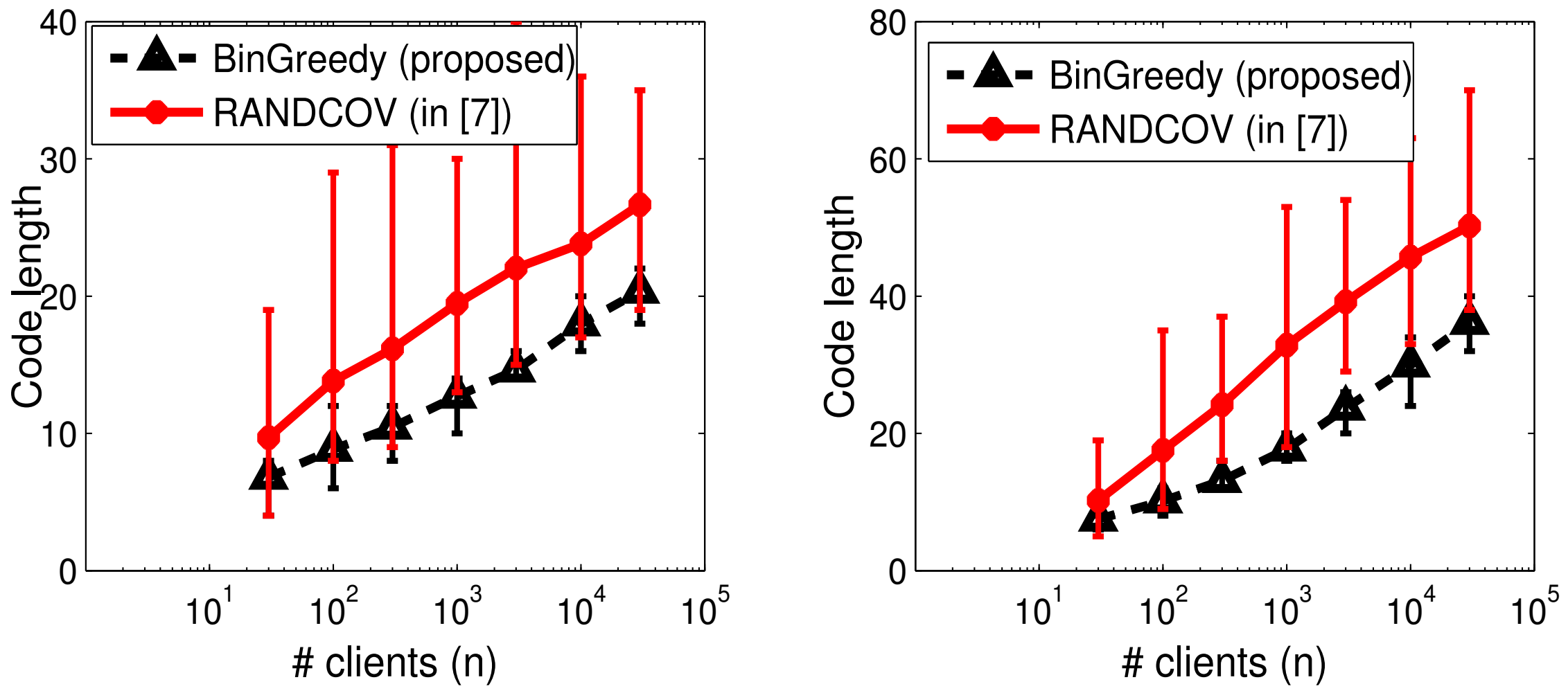

(a) Homogeneous case (b) Heterogeneous case

Fig. 4: Comparison of BinGreedy and randomized algorithms (code length vs. the number of clients). The curves in the figures show the average performance over random instances and the bars at each point show the region between the best and worst case performances.

## B. Optimality Gap

We compare our BinGreedy performance with the optimal binary code length calculated through the minrank method in Section VIII. By setting $n = 12$ and $18$, we evaluate the performance of the two algorithms as $m$ varies[5]. For each pair of $m$ and $n$, we randomly generate $10$ bipartite graph instances by connecting each client and each message with probability $0.3$.

The *gap* for an instance $I$ is defined as the difference of code length achieved by our BinGreedy algorithm and by the optimal binary algorithm, i.e., $gap = BinGreedy(I) - OPT_2(I)$. We plot the *average gap* and the *maximum gap* among instances generated with the same parameters $m$ and $n$. Fig. 5 shows that the average gap (the black bar) is around $2$ for both $n$=12 and $n$=18; the maximum gap (the white bar) is $3$ for both $n$=12 and $n$=18; the same as the average code lengths achieved by the BinGreedy and optimal algorithms. We also note that the approximation ratio for $n$=18 (2.01) is slightly greater than that for $n$=12 (1.87). In fact, the approximate ratio is known to be no less than $\Omega(\log\log(n))$ from [2], so it grows as $n$ increases.

[5]Because of the exponential complexity of finding the optimal performance we can compare only for small instances.

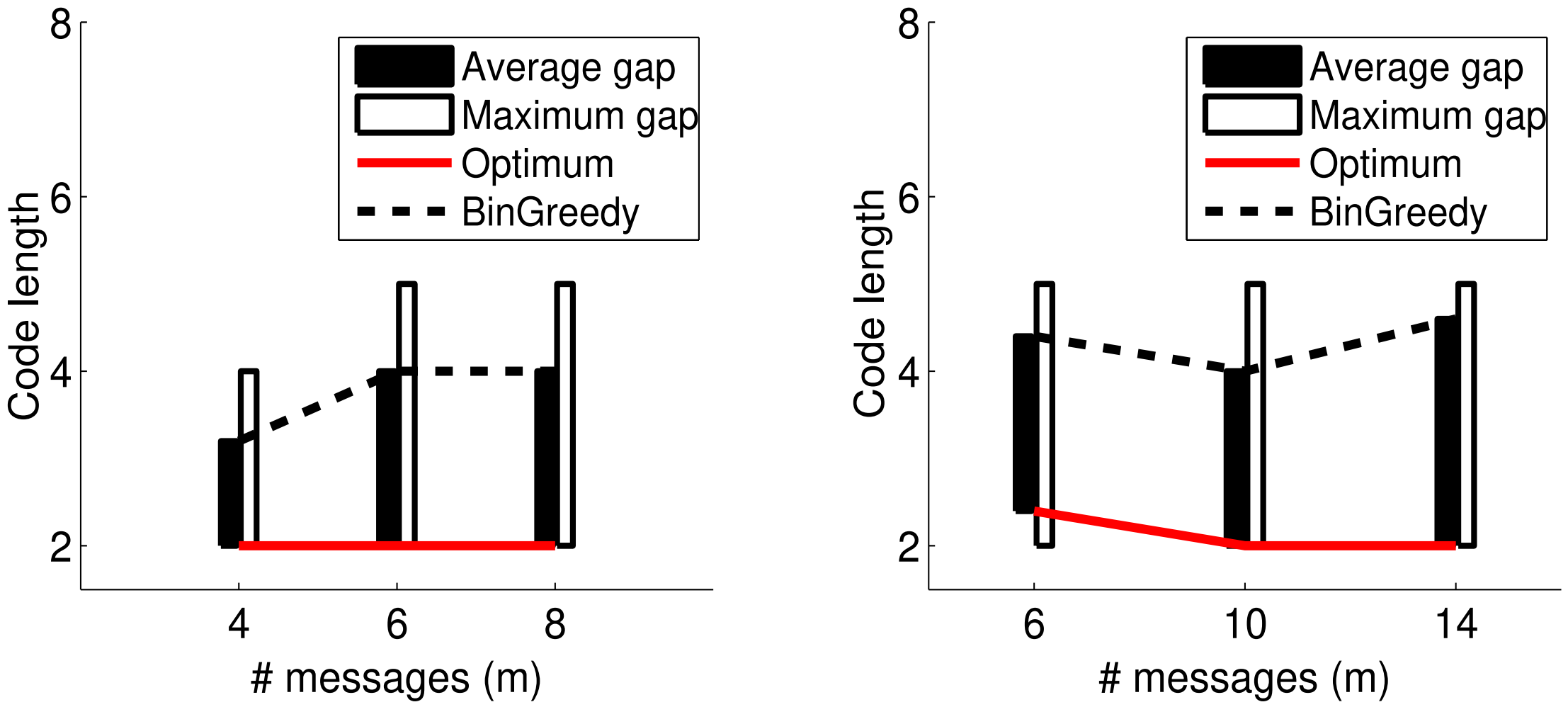

(a) Number of clients $n = 12$ (b) Number of clients $n = 18$

Fig. 5: Optimality gap of BinGreedy algorithm.

### C. $t$-requests Case

In this subsection, we conduct experiments on the $t$-requests case using our BinGreedyT algorithm.

In our simulations, we set the number of messages $m$ to be $n^{0.75}$. We randomly generate $100$ pliable index coding bipartite graph instances for each $n$, by connecting each client and each message with probability $0.3$.

In Fig. 6 (a), we investigate how the code length changes with the number of clients $n$ for $5$-requests and $10$-requests cases. We can see that for both curves, the required code length increases slightly greater than logarithmically with $n$ (notice that the horizontal axis is in logarithmic scale), from $28$ to $42$ for $t = 5$ and from $50$ to $69$ for $t = 10$. Indeed, we show that our algorithm performs in the worst case as $\mathcal{O}(t \log(n) + \log^2(n))$. Given a fixed $t$, we also observe that as $n$ increases, the difference between code lengths in the best case and in the worst case decreases, i.e., the bar in the figure becomes shorter. This implies robustness for larger $n$.

In Fig. 6 (b), we evaluate how the number of requests $t$ affect the the code length for $n = 3000$ and $n = 10000$. We can see that given a fixed number of clients $n$, the code length increases almost linearly with the number of requests $t$, from around $20$ to $60$.

## XI. Conclusions

We have proposed a deterministic polynomial-time algorithm for pliable index coding that achieves code length at most $\mathcal{O}(\log^2(n))$. We modified this algorithm for the $t$-requests case and provided a worst case performance $\mathcal{O}(t\log(n) + \log^2(n))$ guarantee. We constructed problem instances that achieve a lower bound of $\Omega(\log(n))$ for pliable index coding and $\Omega(t+\log(n))$ for the $t$-requests case. However, it is still an open question for the gap between upper and lower bounds, namely, whether we can find some algorithms that achieve an optimal pliable index coding with $\mathcal{O}(\log(n))$ number of transmissions or we can find some instances that require at least $\Omega \log^2(n)$ number of transmissions. We performed a probabilistic analysis over random graphs to show that the optimal code length is almost surely $\Theta(\log(n))$. We also presented experimental results that show up to $50\%$ performance benefits of our proposed algorithms and higher robustness over existing algorithms.

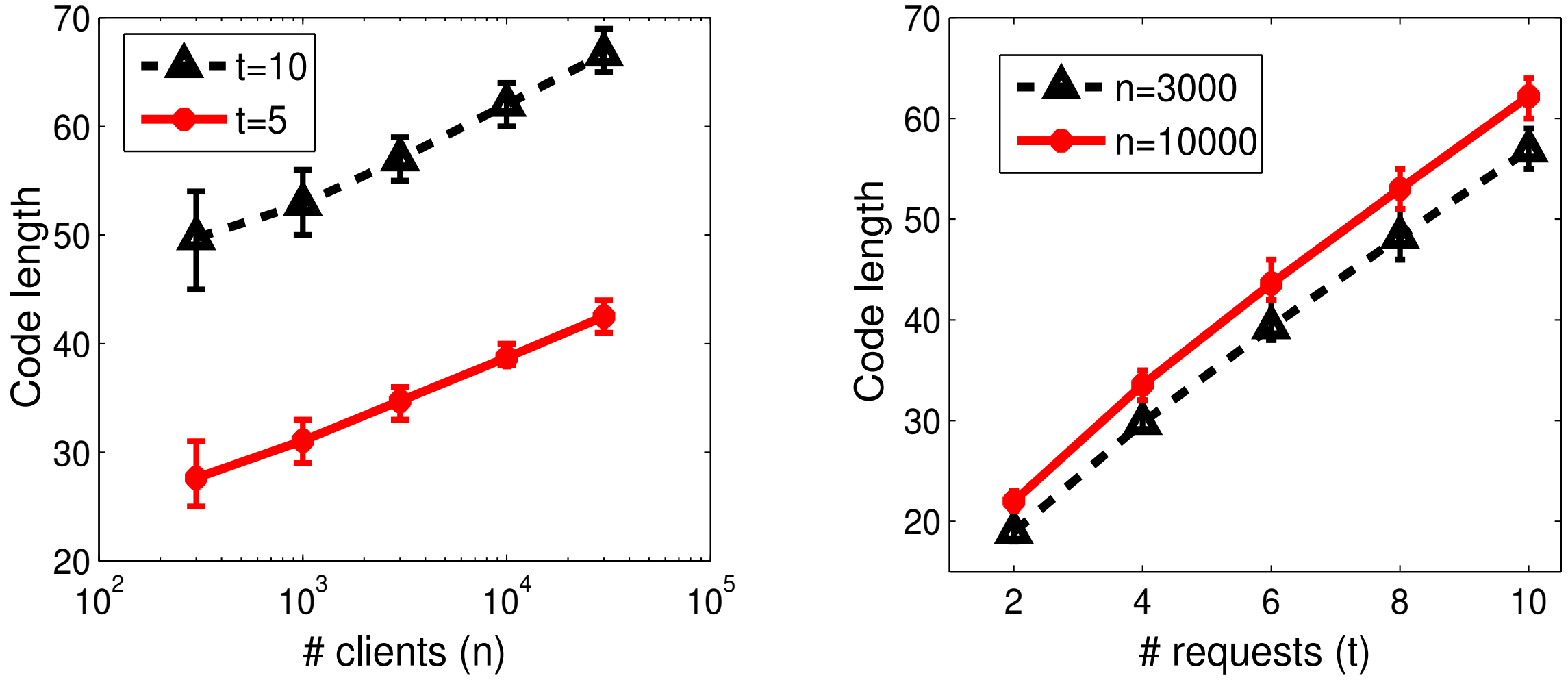

(a) Code length vs. the number of clients (n). (b) Code length vs. the number of requests (t).

Fig. 6: Performance of BinGreedyT algorithm. The curves in the figures show the average performance over random instances and the bars at each point show the region between the best and worst case performances.

## APPENDIX A
## PROOF OF A CLAIM FOR THEOREM 2

We first describe the assignment procedure in detail. Initially, both $SAT$ and $UNSAT$ are empty. We gradually add clients from $\mathcal{N}_s$ into these two sets as we go through the messages and assign coding sub-vectors. Our first step is to add all $N^\dagger[1^{(s)}]$ (the effective neighbors of the first message in $\mathcal{M}_s$) to $SAT$, since any non-zero vector satisfies the decoding criterion for only one message. At each step, some additional clients may become satisfied, but some satisfied clients may also become unsatisfied.

Assume the effective weight of a message $j$ in $\mathcal{M}_s$ is between $w^{(s)}/2$ and $w^{(s)}$. We will show that at each step, the weight of clients who are moved from $SAT$ to $UNSAT$ is at most $w^{(s)}/3$. Consider the step for assigning coding sub-vector to message $j$. Notice that when we assign a coding sub-vector $(1,0)^T$, $(0,1)^T$, or $(1,1)^T$ to message $j$, only clients connecting with message $j$ can be affected. We list possibilities for all the clients, with total weight $h$, that are connected with $j$ and satisfied (in $SAT$) at the beginning of this step:

- Case 1: Assume there are weight $h_1$ worth of clients who connect with previously visited messages that are assigned one coding sub-vector $(1,0)^T$ and some (perhaps none) coding sub-vectors $(0,1)^T$. In this case, these clients can decode a new message corresponding to the coding sub-vector $(1,0)^T$ since $(1,0)^T$ does not belong in the span of $(0,1)^T$. Similarly,
- Case 2: weight $h_2$ worth of clients are satisfied by a $(1,0)^T$, several $(1,1)^T$.
- Case 3: weight $h_3$ worth of clients are satisfied by a $(0,1)^T$, several $(1,0)^T$.
- Case 4: weight $h_4$ worth of clients are satisfied by a $(0,1)^T$, several $(1,1)^T$.
- Case 5: weight $h_5$ worth of clients are satisfied by a $(1,1)^T$, several $(0,1)^T$.
- Case 6: weight $h_6$ worth of clients are satisfied by a $(1,1)^T$, several $(1,0)^T$.

When we assign a coding sub-vector $(1,0)^T$ to message $j$, the $h_3+h_6$ worth of clients can still be satisfied according to the decoding criterion. Similarly, if we assign a coding sub-vector $(0,1)^T$ or $(1,1)^T$ to message $j$, then the weight $h_1+h_5$ or $h_2+h_4$ of clients can still be satisfied.

Note that $h_1+h_2+h_3+h_4+h_5+h_6 \geq h$ as there may be overlap among the 6 different cases (e.g., a client is satisfied by one $(1,0)^T$ and one $(0,1)^T$, so she is counted twice in both Case 1 and Case 3). Hence, at least one of $h_3+h_6$, $h_1+h_5$, $h_2+h_4$ should be no less than $h/3$; our greedy algorithm will move at most $2h/3$ worth of clients from $SAT$ to $UNSAT$.

According to the property of our sorting and grouping phases in eq. (10), the weight of $j$'s neighbors who are connected with previously visited messages (and hence are not $j$'s effective clients) is at most $w^{(s)} - w^{\dagger}[j] < w^{(s)}/2$; otherwise, $j$ will be grouped into another group with index smaller than $s$, since $j$'s effective weight would be larger than $w^{\dagger}[j] + w^{(s)}/2 > w^{(s)}$ when performing the sorting process. Furthermore, every $j$'s neighbor in the set $SAT$ is one of these neighbors. Therefore, $h < w^{(s)} - w^{\dagger}[j] < w^{(s)}/2$. Thus, the weight of clients being moved from $SAT$ to $UNSAT$ at current step is at most $2h/3 < 2w^{(s)}/2/3 = w^{(s)}/3$.

# Appendix B

## Proofs of Corollaries for Vector Pliable Index Coding

In this appendix, we will prove the three corollaries for the vector pliable index coding. The main techniques and outline of the proofs are similar to those used for the scalar pliable index coding. For completeness, we will reiterate each corollary and give the detailed proof.

**Corollary 1.** *In a complete instance $(m, n, \{R_i\}_{i\in[n]})$ for vector pliable index coding with vector length $L$, the optimal number of equivalent transmissions is $\Omega(\log(n))$.*

*Proof.* We follow the proof outline of the Theorem 3 using the mathematical induction method. Obviously, we can trivially satisfy all clients with $m = \log(n)$ equivalent transmissions. We will prove that the rank of the coding matrix $\boldsymbol{A}^{\ddagger}$ needs to be at least $mL$ to satisfy all clients. Let $J \subseteq [m]$ denote a subset of message vector indices; for the complete instance, Lemma 6 needs to hold for any subset $J \subseteq [m]$. We denote by $\boldsymbol{A}^{\ddagger}_J$ the submatrix of $\boldsymbol{A}^{\ddagger}$ with $|J|L$ columns corresponding to the message vectors indexed by $J$.

• For $|J| = 1$, to satisfy the clients who miss only one message, $L$ columns of the coding matrix $\boldsymbol{A}^{\ddagger}$ corresponding to a message vector $\boldsymbol{b}_j$, for any $j \in [m]$, should have a full rank $L$. Otherwise, if for example, $L$ columns corresponding to $j_1$ is not full rank, then the client who only requests message $\boldsymbol{b}_{j_1}$ cannot be satisfied according to the condition 1) in Lemma 6. So rank$(\boldsymbol{A}^{\ddagger}_J) = L$ for $|J| = 1$.

• Similarly, for $|J| = 2$, any two $L$ columns of the coding matrix $\boldsymbol{A}^{\ddagger}$ corresponding to two message vectors $\boldsymbol{b}_{j_1}$ and $\boldsymbol{b}_{j_2}$ should have a full rank $2L$. Otherwise, if for example, $2L$ columns corresponding to $j_1$ and $j_2$ do not have a full rank, then some non-zero vector $\boldsymbol{v}$ is in the space span$\{\boldsymbol{a}_{j_1,1}, \boldsymbol{a}_{j_1,2}, \dots, \boldsymbol{a}_{j_1,L}\}$ and also in span$\{\boldsymbol{a}_{j_2,1}, \boldsymbol{a}_{j_2,2}, \dots, \boldsymbol{a}_{j_2,L}\}$. Then, the client who only misses messages $\boldsymbol{b}_{j_1}$ and $\boldsymbol{b}_{j_2}$ cannot be satisfied according to the condition 2) in Lemma 6. So rank$(\boldsymbol{A}^{\ddagger}_J) = 2L$.

• Suppose we have rank$(\boldsymbol{A}^{\ddagger}_J) = \kappa L$ for all $|J| = \kappa$. For $|J| = \kappa + 1$, we can see that if all clients who only miss $\kappa + 1$ messages can be satisfied, then for some $j \in J$, we have $\boldsymbol{v} \notin$ span$\{\boldsymbol{A}^{\ddagger}_{J\setminus\{j\}}\}$ and rank$(\boldsymbol{A}^{\ddagger}_{\{j\}}) = L$ according to Lemma 6 for any $\boldsymbol{v} \in$ span$\{\boldsymbol{A}^{\ddagger}_{\{j\}}\}$. Therefore, rank$(\boldsymbol{A}^{\ddagger}_J) =$ rank$(\boldsymbol{A}^{\ddagger}_{\{j\}}) +$ rank$(\boldsymbol{A}^{\ddagger}_{J\setminus\{j\}}) = (\kappa + 1)L$.

Therefore, to satisfy all the clients, the rank of the coding matrix $\boldsymbol{A}^{\ddagger}$ is $mL$, resulting in the equivalent number of transmissions $K \geq m$, from which the result follows. □

**Corollary 2.** *In a complete instance $(m, n, \{R_i\}_{i\in[n]}, t)$ for vector pliable index coding with vector length $L$, the optimal number of equivalent transmissions is $\Omega(t + \log(n))$.*

*Proof.* Clearly, $m = \log(n+1)+t-1$ equivalent transmissions are enough to satisfy all clients. So we only show at least $\Omega(t + \log(n))$ equivalent transmissions are needed.

By abuse of notation, let us denote by $1^{(1)}, 2^{(1)}, \dots, m_1^{(1)}$ and $1^{(2)}, 2^{(2)}, \dots, m_2^{(2)}$ the indices of Type-1 and Type-2 message vectors and by $[1^{(1)} : m_1^{(1)}]$ and $[1^{(2)} : m_2^{(2)}]$ the sets of these two types of message vectors.

Suppose the coding matrix for a $t$-requests case problem is $\boldsymbol{A}^{\ddagger}$. We denote by $\boldsymbol{A}^{\ddagger}_{J\cup[1^{(2)}:m_2^{(2)}]}$ the submatrix of $\boldsymbol{A}^{\ddagger}$ consisting of columns corresponding to message vectors indexed by $J \cup [1^{(2)} : m_2^{(2)}]$, where $J \subseteq [1^{(1)} : m_1^{(1)}]$ is a subset of indices of Type-1 message vectors. We will use induction to prove that the rank of the coding matrix $\boldsymbol{A}^{\ddagger}$ needs to be at least $mL$ for all the clients to be $t$-satisfied according to the decoding criterion. In the complete instance, the decoding criterion needs to hold for all clients, or for all $|J| = 1, 2, \dots, m_1$.

For $J \subseteq [1^{(1)} : m_1^{(1)}]$ and $|J| = 1$, i.e., to satisfy the clients who miss only one Type-1 message, we need rank$(\boldsymbol{A}^{\ddagger}_{J\cup[1^{(2)}:m_2^{(2)}]}) = tL$. Since otherwise, for example if the $L$ columns corresponding to $j_1 \in [1^{(1)} : m_1^{(1)}]$ and all $(t-1)L$ columns corresponding to $[1^{(2)} : m_2^{(2)}]$ have rank less than $tL$, then the clients who requests message vectors $\{j_1\} \cup [1^{(2)} : m_2^{(2)}]$ cannot be $t$-satisfied according to the decoding criterion (cannot decode all $tL$ sub-messages). So rank$(\boldsymbol{A}^{\ddagger}_{J\cup[1^{(2)}:m_2^{(2)}]}) = tL$ for all $|J| = 1$.

Assume we have rank$(\boldsymbol{A}^{\ddagger}_{J\cup[1^{(2)}:m_2^{(2)}]}) = (\kappa + t - 1)L$ for all $J \subseteq [1^{(1)} : m_1^{(1)}]$ with $|J| = \kappa$. For $J \subseteq [1^{(1)} : m_1^{(1)}]$ and $|J| = \kappa + 1$, we can see that according to the induction hypothesis, rank$(\boldsymbol{A}^{\ddagger}_{J\cup[1^{(2)}:m_2^{(2)}]}) \geq (\kappa + t - 1)L$. If rank$(\boldsymbol{A}^{\ddagger}_{J\cup[1^{(2)}:m_2^{(2)}]}) < (\kappa + t)L$, then for any message vector $j \in J$, we can find a column corresponding to $\boldsymbol{a}_{j,l}$ for some $l \in [L]$, such that $\boldsymbol{a}_{j,l} \in$ span$\{\boldsymbol{a}^{j',l'} | j' \in J \cup [1^{(2)} : m_2^{(2)}], l' \in [L], (j', l') \neq (j, l)\}$, since otherwise, columns corresponding to $J \cup [1^{(2)} : m_2^{(2)}] \backslash \{j\}$ are linearly independent and adding all $a_{j,l}$, $\forall l \in [L]$ will give a rank $(\kappa + t)L$ for the submatrix $\boldsymbol{A}^{\ddagger}_{J\cup[1^{(2)}:m_2^{(2)}]}$. Hence, $\boldsymbol{b}_j$ (for any $j \in J$) cannot be decoded by the client who is only connected with $J \cup [1^{(2)} : m_2^{(2)}]$. This client can decode at most $t-1$ messages and cannot be $t$-satisfied. As a result, rank$(\boldsymbol{A}^{\ddagger}_{J\cup[1^{(2)}:m_2^{(2)}]}) = (\kappa + t)L$, from which the result follows. □

**Corollary 3.** *For vector pliable index coding with vector length $L$ over random graph $B(m, n, p)$ ($m = \mathcal{O}(n^{\delta})$ for some constant $\delta$), with probability at least $1 - \mathcal{O}(1/n^2)$, the equivalent linear* *pliable index code length can be lower bounded as follows:*

$$K \geq \begin{cases} \frac{\log(n)}{4\log(1/p)}, & p \leq \frac{\sqrt{5}-1}{2}, \\ \frac{\log(n)}{2\log[1/(1-p)]}, & p > \frac{\sqrt{5}-1}{2}. \end{cases} \tag{27}$$

*Proof.* This proof follows the proof outline of Theorem 5 by extending a vector in the coding matrix for scalar pliable index coding to a linear subspace in the vector pliable index coding. According to linear algebra, a set of linear subspaces $\{W_1, W_2, \ldots, W_\mu\}$ are linearly independent if and only if for any subspace $W_j$, $\forall j \in [\mu]$, any vector $\boldsymbol{v} \in W_j$ is not in the span of the sum of the other subspaces $\sum_{j' \in [\mu] \setminus \{j\}} W_{j'}$[6].

For a coding matrix $\boldsymbol{A}^{\ddagger} \in \mathbf{F}_q^{KL \times m}$, we denote by $W_j$, $\forall j \in [m]$, the linear subspace spanned by columns of $\boldsymbol{A}^{\ddagger}$ corresponding to message vector $j \in [m]$. We next show that it suffices that we only consider the case $\text{rank}\{W_j\} = L$ or 0, $\forall j \in [m]$. Indeed, if a subspace $W_j$ of the matrix $\boldsymbol{A}^{\ddagger}$ has rank $0 < \text{rank}\{W_j\} < L$, any client who misses message vector $j$ cannot decode all sub-messages of message vector $j$ (but may be able to decode parts of the sub-messages) according to the condition 1) of the decoding criterion in Lemma 6. Therefore, by setting $W_j = 0$, we achieve a new coding matrix, denoted by $\boldsymbol{A}_0^{\ddagger}$, and as a result, all message vectors a client $i$ can decode using coding matrix $\boldsymbol{A}^{\ddagger}$ can be decoded using coding matrix $\boldsymbol{A}_0^{\ddagger}$.

Let us denote by $\mathcal{A}^{\ddagger} \subseteq \mathbf{F}_q^{KL \times m}$ the set of coding matrices satisfying $\text{rank}\{W_j\} = L$ or 0, $\forall j \in [m]$ . Similar to Section VII, we define a *coding structure* as $S(J^{(1)}, J^{(2)}, J^{(3)}) \triangleq \{\boldsymbol{A}^{\ddagger} \in \mathcal{A}^{\ddagger} | \boldsymbol{A}^{\ddagger}$ satisfies Properties (4) (5) (6)$\}$, where $J^{(1)}, J^{(2)}, J^{(3)} \subseteq [m]$ are disjoint subsets of message vector indices, $|J^{(1)}| + |J^{(2)}| = K$, $|J^{(2)}| = |J^{(3)}|$, and the properties are listed as follows.

**Property.**

*(4) Linear subspaces indexed by $J^{(1)}$ and $J^{(2)}$ contain a maximum number of independent linear subspaces. In other words, any other linear subspace $W_j$, $\forall j \notin J^{(1)} \cup J^{(2)}$, is not independent of the sum of subspaces indexed by $J^{(1)}$ and $J^{(2)}$.*

*(5) For any $j \in J^{(1)}$, the corresponding subspace $W_j$ is independent of the sum of other subspaces, i.e., $W_j$ is independent of $\sum_{j' \in [m] \setminus \{j\}} W_{j'}$ for all $j \in J^{(1)}$.*

*(6) For any $j \in J^{(2)} \cup J^{(3)}$, the corresponding subspace $W_j$ is not independent of the sum of*

[6]The sum of a set of linear subspaces is the linear space spanned by all basis vectors of these linear subspaces.

*other subspaces indexed by* $J^{(2)} \cup J^{(3)}$*, i.e.,* $W_j$ *is not independent of* $\sum_{j' \in J^{(2)} \cup J^{(3)} \setminus \{j\}} W_{j'}$ *for all* $j \in J^{(2)} \cup J^{(3)}$.

We next describe a similar procedure as in the scalar pliable index coding case that maps a given coding matrix $\boldsymbol{A}^{\ddagger} \in \mathcal{A}^{\ddagger}$ (in a non-unique way) to some coding structure $S(J^{(1)}, J^{(2)}, J^{(3)})$. Additionally, given three disjoint subsets of message vector indices $J^{(1)}, J^{(2)}, J^{(3)} \subseteq [m]$, $|J^{(1)}| + |J^{(2)}| = K$, $|J^{(2)}| = |J^{(3)}|$, we can easily find some matrix $\boldsymbol{A}^{\ddagger}$ that satisfies Properties (4), (5), and (6). Thus, if we denote the union of all coding structures by $\mathcal{S} = \cup S(J^{(1)}, J^{(2)}, J^{(3)})$, it is easy to see that $\mathcal{S} = \mathcal{A}^{\ddagger}$.

*Mapping Procedure:* In the following, we will call the subspaces in $J^{(1)}$, $J^{(2)}$ and $J^{(3)}$ Type-1, Type-2 and Type-3 subspaces, respectively. We will use the notation $K_1 = |J^{(1)}|$, $K_2 = |J^{(2)}|$, $K_3 = |J^{(3)}|$. We will show that we can select $J^{(1)}$, $J^{(2)}$ and $J^{(3)}$ so that $K_1 + K_2 = K$, $K_2 = K_3$ and properties (4)-(6) are satisfied. Note that a matrix $\boldsymbol{A}^{\ddagger}$ can be mapped to multiple structures, since there may exist different choices for selecting the columns in $J^{(1)}$, $J^{(2)}$ and $J^{(3)}$.

• In the coding matrix $\boldsymbol{A}^{\ddagger}$, find an arbitrary maximum number of linearly independent subspaces. There are at most $K$ such subspaces and without loss of generality, we assume these subspaces are indexed by $1, 2, \ldots, K'$, where $K' \leq K$.

• We categorize all $m$ subspaces into 3 groups: 2 groups for these $K'$ maximum independent subspaces and a third group for the remaining $m - K'$ subspaces.

– Group 1: $\boldsymbol{W}^{\ddagger(1)} = \{W_{j_1} | W_{j_1}$ is independent of $\sum_{j \in [m] \setminus \{j_1\}} W_j, \forall j_1 \in [K']\}$. Group 1 consists of subspaces that are independent of the sum of other subspaces. We assume $K_1 \leq K'$ such subspaces, and without loss of generality, we assume these subspaces in Group 1 are indexed by $1, 2, \ldots, K_1$. These are the Type-1 subspaces.

– Group 2: $\boldsymbol{W}^{\ddagger(2)} = \{W_{j_2} | W_{j_2}$ is not independent of $\sum_{j \in [m] \setminus \{j_2\}} W_j, \forall j_2 \in [K']\}$. Group 2 consists of subspaces that are not independent of the sum of other subspaces, i.e., among those $K'$ subspaces that are not in Group 1. We assume $K_2 = K' - K_1$ such subspaces, and these subspaces in Group 2 are indexed by $K_1 + 1, K_1 + 2, \ldots, K_1 + K_2 = K'$. These are the Type-2 subspaces.

– Group 3: $\boldsymbol{W}^{\ddagger(3)} = \{W_{j_3} | j_3 \notin [K']\}$. Group 3 consists of the remaining $m - K'$ subspaces.

• We select and label $K_3$ subspaces in Group 3 as Type-3 subspaces as follows. Initially, we mark all $K_2$ subspaces in Group 2 as *active* and we will repeatedly *deactivate* them in the following steps.

1) We pick an arbitrary non-zero subspace $W_{j_3}$ from Group 3 that was not picked before.

2) Label subspaces or discard them according to the following rule.

Recall that $\boldsymbol{a}_{K_1+1,1}, \boldsymbol{a}_{K_1+1,2}, \ldots, \boldsymbol{a}_{K_1+1,L}; \boldsymbol{a}_{K_1+2,1}, \boldsymbol{a}_{K_1+2,2}, \ldots, \boldsymbol{a}_{K_1+2,L}; \ldots; \boldsymbol{a}_{K_1+K_2,1}, \boldsymbol{a}_{K_1+K_2,2}, \ldots, \boldsymbol{a}_{K_1+K_2,L}$ are the bases of the $K_2$ subspaces in Group 2 and they together form a basis for the sum of the $K_2$ subspaces in Group 2. For short, we call the basis of the sum of the $K_2$ subspaces in Group 2 the *basis of Group 2*. Also note that $\boldsymbol{a}_{j_3,1}, \boldsymbol{a}_{j_3,2}, \ldots, \boldsymbol{a}_{j_3,L}$ is a basis of the subspace $W_{j_3}$.

We check all basis vectors $\boldsymbol{a}_{j_3,1}, \boldsymbol{a}_{j_3,2}, \ldots, \boldsymbol{a}_{j_3,L}$ for subspace $W_{j_3}$. If $\boldsymbol{a}_{j_3,l_3}$, $\forall l_3 \in [L]$ can be represented as a linear combination of the basis of Group 2: $\boldsymbol{a}_{j_3,l_3} = \lambda^{l_3}_{K_1+1,1}\boldsymbol{a}_{K_1+1,1} + \ldots + \lambda^{l_3}_{K_1+1,L}\boldsymbol{a}_{K_1+1,L} + \ldots + \lambda^{l_3}_{K_1+K_2,1}\boldsymbol{a}_{K_1+K_2,1} + \ldots + \lambda^{l_3}_{K_1+K_2,L}\boldsymbol{a}_{K_1+K_2,L}$. According to linear algebra, either we cannot express or we can express uniquely $\boldsymbol{a}_{j_3,l_3}$ as a linear combination of the basis of Group 2. According to the above grouping procedure, we can express at least one of the basis vectors of subspace $W_{j_3}$ as a linear combinations of the basis Group 2. Otherwise, the selected $K'$ subspaces are not the maximum independent ones, as we can add $W_{j_3}$ into them.

For those $\boldsymbol{a}_{j_3,l_3}$ that can be expressed as a linear combination of basis of Group 2, we can consider $(\lambda^{l_3}_{K_1+1,1}, \ldots, \lambda^{l_3}_{K_1+1,L}, \ldots, \lambda^{l_3}_{K_1+K_2,1}, \ldots, \lambda^{l_3}_{K_1+K_2,L})$ as coordinates under this basis. We consider the subspaces in Group 2 that contribute to the non-zero coordinates, i.e., $\boldsymbol{W}^{\ddagger(2)}_0 = \{W_{j_2} \in \boldsymbol{W}^{\ddagger(2)} | \lambda^{l_3}_{j_2,l_2} \neq 0, \exists l_2, l_3 \in [L]\}$. If no subspaces in $\boldsymbol{W}^{\ddagger(2)}_0$ are marked *active*, then discard subspace $W_{j_3}$ without labeling it. If any of these subspaces in $\boldsymbol{W}^{\ddagger(2)}_0$ is marked *active*, then label the subspace $j_3$ as a Type-3 subspace, and mark all subspaces in $\boldsymbol{W}^{\ddagger(2)}_0$ as *inactive* if they are still *active*.

3) Repeat Steps 1) and 2) until all vectors in Group 2 are marked *inactive*. This can always be achieved. Indeed, according to the definition of Group 2, any subspaces $W_{j_2} \in \boldsymbol{W}^{\ddagger(2)}$ is not independent of the sum of other subspaces. So, some vector $\boldsymbol{a}_{j_2,l_2}$ can always appear as a non-zero coordinate in the linear expansion for some subspace in Group 3; otherwise $W_{j_2}$ belongs to Group 1.

We observe that after the above process, there are $K_1$ Type-1 subspaces, $K_2$ Type-2 subspaces, and at most $K_2$ Type-3 subspaces. This is because when we label each Type-3 subspace, we always set *inactive* at least 1 subspace in Group 2.

To deal with the case that $K'$ is less than $K$, we arbitrarily label $K - K'$ unlabeled subspaces in Group 3 as Type-2 subspaces to make $K_1 + K_2 = K$; we can also arbitrarily mark another $K_2 - K_3$ unlabeled subspaces in Group $3$ as Type-3 subspaces to make $K_2 = K_3$. It is easy to see that after this padding, the selected Type-1, Type-2, and Type-3 subspaces satisfy the desired properties.

Note that if a client $i$ has the following connection pattern: $j' \notin R_i$ for any column $j' \in J^{(1)}$ and $j'' \in R_i$ for any column $j'' \in J^{(2)} \cup J^{(3)}$, then client $i$ cannot be satisfied by the coding matrices in coding structure $S$ according to our decoding criterion in Lemma 6. The corollary then follows from the same arguments as in the scalar case in Lemmas 4, 5 and Theorem 5. $\square$